\documentclass[preprint,3p]{elsarticle}

\usepackage{amssymb}
\PassOptionsToPackage{hyphens}{url}\usepackage{hyperref}

\usepackage[T1]{fontenc}
\usepackage[utf8]{inputenc}
\usepackage[english]{babel}
\usepackage{soul,color}
\usepackage{booktabs}
\usepackage[protrusion=true,expansion=true]{microtype}	
\usepackage{amsmath,amsfonts,amsthm} 
\usepackage{graphicx}	
\usepackage{lastpage}
\usepackage{subcaption}
\usepackage{algorithm}
\usepackage{algpseudocode}

\usepackage{txfonts}
\usepackage{lscape}
\usepackage{multicol}
\usepackage{algcompatible}
\usepackage[shortlabels]{enumitem}
\usepackage{multirow}
\usepackage{array}
\newcolumntype{P}[1]{>{\centering\arraybackslash}p{#1}}
\usepackage[table]{xcolor}

\algnewcommand\algorithmicinput{\textbf{Input:}}
\algnewcommand\INPUT{\item[\algorithmicinput]}

\algnewcommand\algorithmicoutput{\textbf{Output:}}
\algnewcommand\OUTPUT{\item[\algorithmicoutput]}
\usepackage{pifont}

\algdef{SE}[PROCEDURE]{Procedure}{EndProcedure}%
[2]{\algorithmicprocedure\ \textproc{#1}\ifthenelse{\equal{#2}{}}{}{(#2)}}%
{\algorithmicend\ \algorithmicprocedure}%

\usepackage[utf8]{inputenc}

\usepackage{mathtools}
\usepackage[figuresright]{rotating}
\algnewcommand{\Initialize}[1]{%
                                  \State \textbf{Initialize:}
                                  \Statex \hspace*{\algorithmicindent}\parbox[t]{.8\linewidth}{\raggedright #1}
                                }

\newcounter{phase}[algorithm]
\newlength{\phaserulewidth}
\newcommand{\setphaserulewidth}{\setlength{\phaserulewidth}}

\makeatother

\setphaserulewidth{.7pt}

\journal{}

\begin{document}

\begin{frontmatter}

\title{UAV-assisted Internet of Vehicles: A Framework Empowered \\by Reinforcement Learning and Blockchain}

\author[add1]{Ahmed Alagha}
\author[add2,add3]{Maha Kadadha}
\author[add2,add3]{Rabeb Mizouni \corref{cor1}}
\author[add2,add3]{Shakti Singh}
\author[add2,add4,add1]{Jamal Bentahar}
\author[add2,add3]{Hadi Otrok}

\address[add1]{Concordia Institute for Information Systems Engineering, Concordia University, Montreal, QC, Canada}
\address[add2]{Department of Computer Science, Khalifa University, Abu Dhabi, UAE}
\address [add3]{Center of Cyber Physical Systems (C2PS), Khalifa University, Abu Dhabi, UAE}
\address [add4]{6G Research Center, Khalifa University, Abu Dhabi, UAE}

\cortext[cor1]{Corresponding author}

\begin{abstract}
This paper addresses the challenges of selecting relay nodes and coordinating among them in UAV-assisted Internet-of-Vehicles (IoV). Recently, UAVs have gained popularity as relay nodes to complement vehicles in IoV networks due to their ability to extend coverage through unbounded movement and superior communication capabilities. The selection of UAV relay nodes in IoV employs mechanisms executed either at centralized servers or decentralized nodes, which have two main limitations: 1) the traceability of the selection mechanism execution and 2) the coordination among the selected UAVs, which is currently offered in a centralized manner and is not coupled with the relay selection. Existing UAV coordination methods often rely on optimization methods, which are not adaptable to different environment complexities, or on centralized deep reinforcement learning, which lacks scalability in multi-UAV settings. Overall, there is a need for a comprehensive framework where relay selection and coordination processes are coupled and executed in a transparent and trusted manner. This work proposes a framework empowered by reinforcement learning and Blockchain for UAV-assisted IoV networks. It consists of three main components: a two-sided UAV relay selection mechanism for UAV-assisted IoV, a decentralized Multi-Agent Deep Reinforcement Learning (MDRL) model for efficient and autonomous UAV coordination, and finally, a Blockchain implementation for transparency and traceability in the interactions between vehicles and UAVs. The relay selection considers the two-sided preferences of vehicles and UAVs based on the Quality-of-UAV (QoU) and the Quality-of-Vehicle (QoV). Upon selection of relay UAVs, the coordination between the selected UAVs is enabled through an MDRL model trained to control their mobility and maintain the network coverage and connectivity using Proximal Policy Optimization (PPO). MDRL offers decentralized control and intelligent decision-making for the UAVs to maintain coverage and connectivity over the assigned vehicles. The evaluation results demonstrate that the proposed selection mechanism improves the stability of the selected relays, while MDRL maximizes the coverage and connectivity achieved by the UAVs. Both methods show superior performance compared to several benchmarks.


\end{abstract}

\end{frontmatter}

\section{Introduction}
\label{Sec: Intro}

The Internet of Vehicles (IoV) field has bloomed in recent years with the rise of smart cities. IoV integrates concepts from Vehicular Ad Hoc Networks (VANETs) and Internet of Things (IoT), where Internet-enabled vehicles equipped with sensors are connected \cite{HEMMATI2023200226, alagha2021sdrs, alagha2020rfls}. Their connectivity paves the way to various use cases, such as reducing traffic jams, transit time, and accidents \cite{krishna2020survey}, facilitating media playback, and file-sharing. Despite its usability, the decentralized nature of IoV makes it challenging to select high-quality and stable relay nodes in the network and to maintain coverage, given the limited number of deployed Roadside Units (RSUs).

Unmanned Aerial Vehicles (UAVs) have been gaining interest in both the fields of VANETs and IoV due to their autonomous flight, unbounded movement patterns, and higher communication capabilities \cite{10.1007/978-981-15-9031-3_14}. They offer deployment flexibility to IoV/VANET networks compared to RSUs, as they can be deployed on the fly. In addition, UAVs improve connectivity, load balancing, and information propagation delay between ground vehicles with superior communication capabilities. UAV-assisted IoV networks have three challenges that existing works attempt to tackle: 1) The \textit{selection of relay nodes} where different mechanisms can be used to assign vehicles to relay UAVs, 2) the traceability of selection mechanism \textit{execution}, and 3) the \textit{coordination} between relay UAVs to maximize connectivity and coverage in the network.


Different mechanisms have been used for the selection of relays in IoV/VANET including optimization-based methods such as quasi-convex optimization \cite{10117548}, auction-based selection \cite{9292475}, and game theoretic approaches such as hedonic, evolutionary, Stackelberg, and matching games  \cite{ABUALOLA2021100355,10036008,ABUALOLA2021100290}. However, these mechanisms are applied either on servers or internally at nodes, which compromises trust and transparency due to the possibility of attacks on servers and malicious behaviour by nodes in the network. Several works proposed the use of Blockchain \cite{nsatoshi} for IoV networks due to its ability to provide a trusted, decentralized, and transparent framework with no-down time compared to centralized systems. In specific, it has been used for securing  transactions between communicating nodes in IoVs, enabling vehicle authentication, storing data on the Blockchain, and managing AI models \cite{KADADHA2021102502,9125437,8911664,10104127,ALLADI2020100249}. While Blockchain improves trust and transparency, it has not been adopted in UAV-assisted IoV networks.

Once selected, UAVs have significant importance in maintaining the connectivity and coverage over vehicles in IoVs. Connectivity here is a measure of how well the UAVs maintain communication links with one another, which requires them to continuously position themselves within each other's communication ranges to ensure a robust and interconnected network. On the other hand, coverage refers to the extent to which UAVs can provide network access to vehicles within their communication ranges. To achieve this, the UAVs need to coordinate and position themselves intelligently to maximize the coverage according to the distribution and mobility of vehicles, without violating connectivity. Some works in the literature address the coordination between selected UAVs by optimizing their placement in the area, such as the coverage and connectivity are maximized \cite{islam2022dynamic, sedjelmaci2019toward}. However, due to the dynamicity of VANET environments, such static strategies may result in sub-optimal performance, loss of connectivity, and deteriorating coverage. Alternatively, recent works focus on continuous and intelligent decision-making, where the UAVs coordinate to move in the environment according to the mobility of the vehicles \cite{raza2021uav, 9082162, 9585312, 9566766}. Most of these works use centralized control stations that organize the positioning and communications of UAVs, which struggle in terms of scalability with the number of agents.

To address the aforementioned drawbacks in relay selection and UAV coordination, this paper proposes a comprehensive management framework for UAV-assisted IoV networks leveraging Blockchain and Multi-Agent Deep Reinforcement Learning (MDRL). The Blockchain-based framework enables two main features: 1) a Blockchain-based UAV selection mechanism for relay selection and 2) decentralized MDRL-based multi-agent coordination to ensure scalable cooperation between the UAVs in maintaining connectivity and coverage. The Blockchain-based UAV selection enables the selection of vehicles to UAVs by considering both of their preferences in a transparent and trusted manner. Vehicles first determine the Quality-of-UAV \textit{(QoU)} for available UAVs within their range and submit the maximum QoU UAV along with their requirements to the smart contract. Then, the smart contract allocates UAVs to proposing vehicles given their bandwidth capacity and the ranked list of vehicles based on their Quality-of-Vehicle \textit{(QoV)}. For coordination between relay UAVs, the problem is formulated as a Markov Game where an MDRL model is trained to decide the mobility actions in the environment in decentralized multi-agent settings to maintain the connectivity between the UAVs and the coverage of the assigned vehicles. A Convolutional Neural Network (CNN) is designed as the actor-network that translates each UAV's observations into mobility actions. A team-based reward function is designed to allow for cooperation between the agents. The learning is optimized using Proximal Policy Optimization (PPO) in a Centralized Learning and Distributed Execution (CLDE) manner. The storage of the trained MDRL models is managed using the InterPlanetary File System (IPFS), which are assigned to UAVs upon selection. 

In summary, the contributions of this work are:

\begin{enumerate}
    \item A transparent and trusted management framework for UAV-assisted IoV through Blockchain.
    \item A Blockchain-hosted two-sided relay selection mechanism incorporating specifically designed QoV and QoU metrics for UAV-assisted IoV.
    \item The formulation of the UAV coordination problem as a Markov Game and the design of the UAV agents' observations in an optimized manner to capture essential information in the environment
    \item The design of a team-based reward function for MDRL that pushes the UAV agents to cooperate to maximize and maintain connectivity and coverage in a dynamic environment.
\end{enumerate}

The proposed selection and coordination methods demonstrate improved performance when compared to existing benchmarks. The selection mechanism shows improved QoU for vehicles and QoV for UAVs. Upon selection, the proposed MDRL method shows improved connectivity and coverage while being scalable to different team sizes (number of UAVs) and environment complexities (number of assigned vehicles). In addition, the implementation of the framework on Solidity demonstrates its feasibility.

\section{Related Work}

This section covers related works in different domains related to the targeted problem, including \textit{IoV Relay Selection}, \textit{Blockchain for Trusted Execution}, and \textit{UAV Coordination in IoV}.
\label{section: Related Work}

\subsection{IoV Relay Selection}

The selection of relay nodes in IoV/ VANET networks is an important challenge due to the impact on the stability of the network, hence the network performance. While only vehicles were considered initially during the selection of relay nodes, UAVs rose as a solution to improve the connectivity and coverage of IoVs. 

Existing works considered different mechanisms for the selection of relay nodes such as optimization, auctions, and game theory. The works in \cite{10117548,9292475,10036008} propose mechanisms that run on edge servers for the formation of the network. In \cite{10117548}, the authors optimize the positions of UAV relays given their driving trajectory. They consider the cooperation between UAVs where the transmission power for each UAV is optimized along with the deployment positions. In \cite{9292475}, an auction mechanism is applied for the selection of UAV coalitions delegated tasks in the IoV to improve communication efficiency. Bids are considered an indication of the preferences of workers and are used to incentivize the participation of UAVs when the selection mechanism forms stable coalitions. In \cite{10036008}, the evolutionary algorithm is utilized for clustering vehicles and placing services on UAVs. The algorithm accounts for the cluster lifetime, the connection time, and the energy of the UAVs. In \cite{ABUALOLA2021100355,ABUALOLA2021100290}, hedonic and matching game models are proposed respectively with the game models running in a decentralized manner at the nodes of the network. In \cite{ABUALOLA2021100355}, the authors propose the use of a hedonic game model to form stable coalitions by accounting for the Quality-of-Service (QoS) of vehicles. In \cite{ABUALOLA2021100290}, they extend their work and consider the existence of UAVs as part of the IoV network. They consider the two-sided preferences of UAVs and vehicles in the matching mechanism for stable selections to be formed.

While different mechanisms have been applied for the selection of nodes, vehicles, and UAVs, the aforementioned work all lack transparency and traceability in the selections formed where they rely on the trust of the servers and nodes. 

\subsection{Blockchain for Trusted Execution}

Multiple domains such as last-mile delivery \cite{KADADHA2024100761,kadadha2022context}, VANET \cite{KADADHA2021102502,9125437,10104127}, and UAVs \cite{ALLADI2020100249} adopted Blockchain as an alternative to traditional centralized systems. Blockchain offers trust and transparency to these domains along with no-down for deployed systems.  In \cite{KADADHA2024100761}, Blockchain is considered in an environment where vehicles and UAVs needed to be selected last-mile delivery tasks. The preferences of both were accounted for in a Blockchain-hosted mechanism offering transparency and trust in the selection. In \cite{8911664}, the different use cases for Blockchain in IoV are defined where the authors focus on the impact of Blockchain on securing the transactions between communicating nodes in the IoV. In \cite{KADADHA2021102502,9125437,10104127}, the authors considered the integration of Blockchain with vehicular networks. In \cite{KADADHA2021102502}, the authors propose a Blockchain-enabled mechanism for the selection of relay nodes in vehicular ad hoc networks in a transparent and trusted manner. Blockchain is utilized to manage vehicles' reputation and transparent storage of the selection mechanism. Meanwhile, the authors \cite{9125437} proposed the use of Blockchain for securing IoV networks by enabling the authentication of vehicles and storing the data on the Blockchain. In \cite{10104127}, the authors propose the division of vehicles and RSUs into different zones where the vehicles register to the Blockchain and RSUs communicate securely through the Blockchain. In \cite{ALLADI2020100249}, the authors tackled security UAV networks by utilizing Blockchain for securing data exchange and communication links. 

Despite the utilization of Blockchain in different domains, it has not yet been adopted in UAV-assisted networks to offer trust and transparency for the selection of relay nodes. Yet, key factors to account for moving forward are the complexity and scalability of the adopted blockchain to the domain. In fact, while centralized systems offer easier management and can handle complex mechanisms, decentralized systems such as blockchain offer better scalability and enhanced privacy. In addition, based on our knowledge, the only work that considers on-chain relay selection is \cite{KADADHA2021102502}, where nodes are vehicles only without considering UAVs. Therefore, key factors to account for moving forward are the complexity and scalability of the adopted Blockchain to the domain.

\subsection{UAV Coordination}
The positioning of the chosen UAVs is an important aspect in UAV-assisted IoVs to ensure good coverage and connectivity. Several works in the literature have addressed this problem using optimization methods. In \cite{8716508}, the authors use Successive Convex Approximation (SCA) to minimize the number of deployed UAVs and optimize their trajectory aiming to maintain connectivity. The authors in \cite{9076813} deploy UAVs in real-time based on predicted vehicular distribution, where a Multimodal Nomad Algorithm (MNA) is used to decide the best service positions in order to enhance the efficiency of drone-assisted VANETs. Another work in \cite{hadiwardoyo2020three} proposes a three-dimensional UAV positioning technique to support communication between vehicles on the ground using Particle Swarm Optimization (PSO) and a Genetic Algorithm (GA). In \cite{9400369, islam2022dynamic}, the authors propose a PSO-based optimization method for the deployment of collaborative UAVs for urban VANETs, where the aim is to optimize the coverage area. A similar work in \cite{9044857} proposes an ellipse clustering algorithm that optimizes the locations of UAV base stations, amongst other factors, to maximize the coverage of ground users. While the aforementioned optimization-based works show promising results, they have drawbacks in terms of adaptability and scalability to different environment complexities. The variations in unknown environments might require the UAVs to explore for some time before making decisions. Additionally, most of these methods centrally decide on the positioning of the UAVs before deployment and lack dynamicity and continuous decision-making which is essential in realistic scenarios.

To overcome the aforementioned drawbacks, recent works resorted to the use of Deep Reinforcement Learning (DRL) as a method to obtain intelligent UAV agents that are capable of continuous decision-making for efficient coverage and connectivity in VANETs. In \cite{8960481}, the authors employ Q-Learning, a Deep Reinforcement Learning (DRL) technique, to navigate an unfamiliar environment and determine the optimal paths for a minimal number of UAVs, ensuring connectivity for ground vehicles traveling at various speeds. Another work in \cite{9082162} uses the Deep Deterministic Policy Gradient (DDPG) DRL method to govern the UAVs’ trajectories with a set of continuous actions. In \cite{9585312}, the authors propose a DRL method considering UAVs' communication ranges, where a central agent (base station) learns how to optimally control and position the UAVs. Similarly, the authors in \cite{9566766} use Double Deep Q-Networks (DDQN) as a DRL algorithm to obtain an agent that is capable of controlling the UAVs for maximized coverage and connectivity. A major common drawback in the works above is the use of a centralized agent that controls all the UAVs, which results in issues with scalability, single point of failure, and collaboration between the UAVs. MDRL is generally used to address such scalability issues, which have been previously tackled in many applications such as path planning \cite{damani2021primal, bayerlein2021multi}, target search \cite{alagha2023multi}, and mobile edge computing \cite{lee2023multi}. In the context of UAV-assisted IoVs, the authors in \cite{jiang2021marl} propose a MDRL solution, but only focus on UAVs deployed in a single/sparse highway with few vehicles, and not in urban areas crowded with vehicles.



\section{Proposed System}
\begin{figure}[th!] 
\centering
\includegraphics[width=0.85\linewidth]{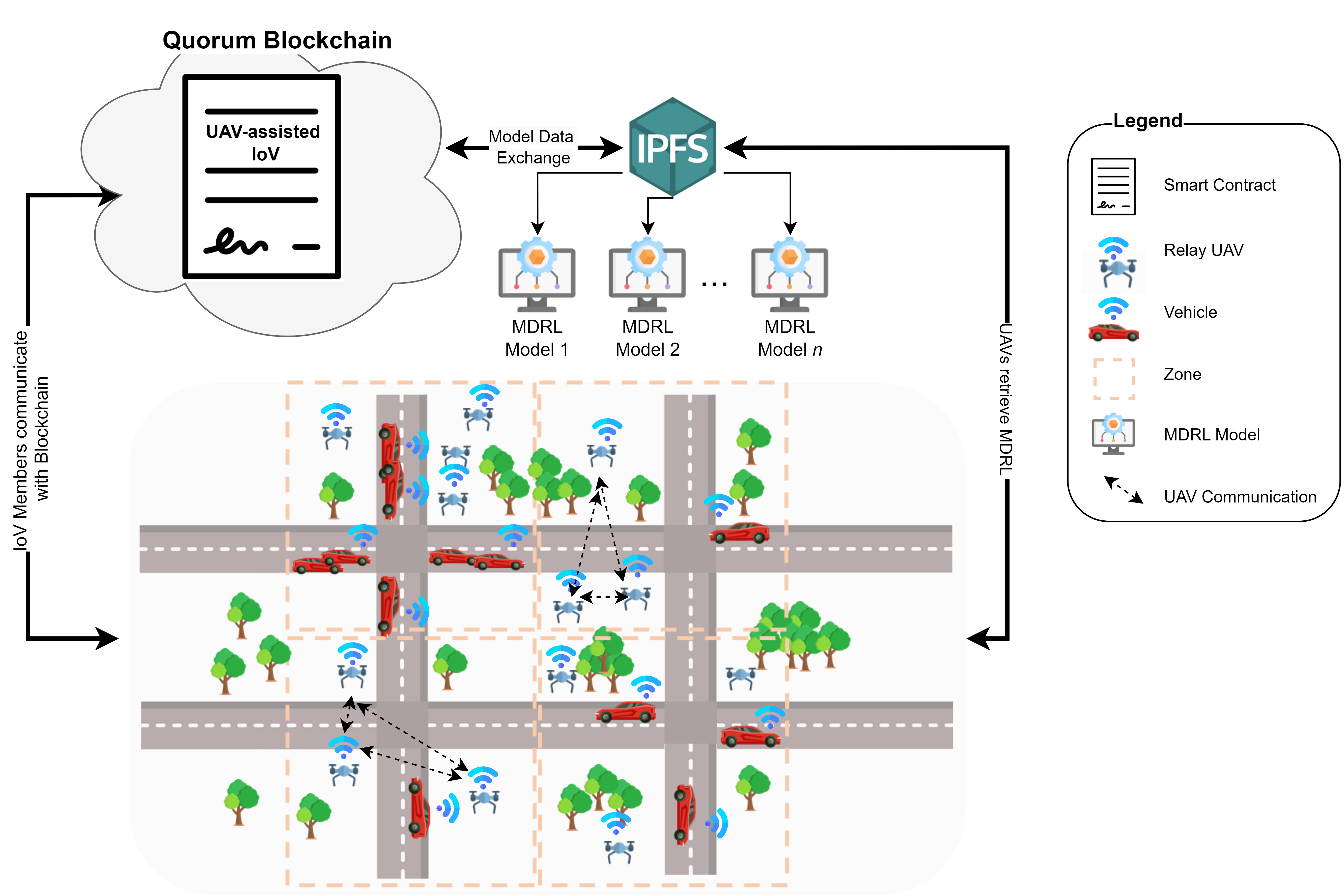}
\caption{The different layers of the system in the proposed framework}
\label{fig:proposedframework}
\end{figure}
This section presents and discusses the different components of the proposed system, including the Blockchain-based framework, UAV-assisted IoV construction, and MDRL-based UAV coordination. An overview of the proposed system is illustrated in Figure \ref{fig:proposedframework} with the general interactions between the system components, where the order of the interactions is illustrated later in Figure \ref{fig:process}. It enables the members of the IoV, being vehicles and relay UAVs, to interact transparently, form a stable network, and maintain connectivity. The physical environment is divided into geographical zones with unique identifiers where vehicles and relay UAVs are assigned to a single zone based on their GPS coordinates. A smart contract is designed for managing the registration of IoV members and the selection of UAVs to vehicles on top of Quorum consortium Blockchain. The proposed MDRL algorithm is used to train several models for different variations of the environment (different number of agents, number of vehicles, etc). Trained MDRL models are stored on the IPFS, which is a decentralized file storage system that efficiently handles large files. In IPFS, the data is distributed across a peer-to-peer network, where files (i.e. models) are replicated across multiple nodes, ensuring they remain accessible even if some nodes fail. In our framework, metadata associated with these models, such as their unique IPFS identifiers (needed to access the file on IPFS), ownership details, and application-specific information about the models are stored on the Blockchain. IPFS is used instead of the Blockchain for storing the actual models due to the Blockchain’s limited storage capacity and the high cost of storing large files. Once UAVs are selected, they are assigned one of the available MDRL models by the smart contracts depending on the environment. Each UAV is given a copy of the model and acts in a decentralized manner based on its observations to maintain connectivity and coverage for the assigned vehicles.



The members of the frameworks and its key components are detailed below.

\begin{itemize}
    \item \textit{IoV members:} which are vehicles and UAVs that form the IoV network. \textit{Vehicles} are interested in maintaining connection to other nodes in the network while \textit{UAVs} are available to extend the connectivity of the network.
    \item \textit{Blockchain-based Framework}: which consists of the smart contract running on the Quorum Blockchain \cite{quorum} that form the UAV-assisted IoV and manages sharing the MDRL model with the UAVs.
\item \textit{UAV-assisted IoV Relay Selection:} which is the mechanism responsible for constructing the IoV by allocating vehicles to UAVs.
    \item \textit{MDRL models for UAV Coordination}: which are the different trained models to control and coordinate the UAVs' mobility in the area, to be assigned to them according to the environment upon selection
    \item \textit{IPFS:} which is used to store trained MDRL models to be assigned and used by the UAVs.
\end{itemize}

\subsection{Proposed Blockchain-based Framework}
The proposed Blockchain-based framework consists of a smart contract that facilitates the interactions between the vehicles and UAVs. Quorum Blockchain \cite{quorum}, which is an Ethereum-based consortium Blockchain, is considered for deploying the designed smart contract to have. Quorum Blockchain was selected for its ability to offer high transaction throughput (up to 3,000 TPS) in reasonable delay \cite{quorum_performance}. The proposed smart contract is responsible for storing and sharing their information, collecting selection proposals from vehicles and allocating them to UAVs, and sharing the adequate MDRL models to UAVs to maintain the stability of the selection.

The designed smart contract in the proposed framework consists of multiple data structures, variables, mappings, and functions. Table \ref{tab:datastructs} presents the three designed data structures being \textit{UAVInfo}, \textit{VehicleInfo}, and \textit{VehicleSelection}. \textit{UAVInfo} maintains the information relevant to a UAV being its location and altitude, reputation, energy level, available bandwidth, and the timestamp for the last information update. It is worth noting that we assume that each UAV hover at the same location  constant prior to the selection process and would only change its location upon selection to maintain connectivity to vehicles selecting it. On the other hand, \textit{VehicleInfo} maintains the information of a vehicle consisting of its location, reputation, the payment its willing to give per Mbps along with the timestamp for the last time the information was updated. \textit{VehicleSelection} saves the information related to a vehicle's determined selection such as its Ethereum address, QoV, requested bandwidth (RB) and the normalized QoV over the square root of the requested bandwidth. The QoV$/\sqrt{RB}$ is considered for ranking the vehicles.
\begin{table}[ht]
\centering
\small
\caption{Data Structures}
\label{tab:datastructs}
\begin{tabular}{|c|c|} 
\hline
\multicolumn{2}{|c|}{\cellcolor{gray!25} \textbf{UAVInfo }} \\\hline \hline
    \centering Location  (string)   &   Altitude (string)\\ \hline
\centering	 Reputation (uint) &  Energy Level (uint) \\ \hline 
Available Bandwidth (uint) & Timestamp (uint)\\
\hline \hline
\multicolumn{2}{|c|}{\cellcolor{gray!25} \textbf{VehicleInfo }} \\\hline \hline
    \centering Location  (string)   &  PayPerMbps (uint)    \\ \hline
\centering	 Reputation (uint) & Timestamp (uint) \\ \hline \hline

\multicolumn{2}{|c|}{\cellcolor{gray!25} \textbf{VehicleSelection}} \\\hline \hline
    \centering Vehicle (address)   & QoV value (uint)  \\ \hline
    Requested Bandwidth (uint) &  $QoV/\sqrt{RB}$ (uint)  \\ \hline
\end{tabular}
\end{table}


Table \ref{tab:variablesandmappings} presents the variables and mappings designed in the smart contract. Three array variables are included: 1) \textit{Vehicle List} which stores the Ethereum addresses of registered vehicles, 2) \textit{UAV List} which stores the Ethereum addresses of registered UAVs, and 3) \textit{Zones} which holds the unique identifiers of zones with IoV members. IoV members are assumed to be able to identify their zones based on their GPS coordinates and a predefined GPS to zone ID conversion. Two mappings are designed to map the address of vehicles and UAVs to their information, \textit{VehicleData List} and \textit{UAVData List} respectively. Meanwhile, \textit{UAVs in zone} maps the Zone ID to the current list of UAVs currently in that zone. Two mappings are designed for the selection mechanism process being \textit{UAV Proposal List} and \textit{UAV Selection List} where they map the UAV address to the list of proposing vehicles and selected vehicles respectively. Lastly, \textit{Selected UAVs} maps the Zone ID to the addresses of selected UAVs, which is required for the coordination through MDRL.
\begin{table}[ht]
\centering
\small
\caption{Variables and Mappings}
\label{tab:variablesandmappings}
\begin{tabular}{|c|c|c|} 
\hline
\multicolumn{3}{|c|}{\cellcolor{gray!25} \textbf{Variables}} \\\hline \hline

Vehicle List (address[]) & UAV List (address[]) & Zones (uint[]) \\ \hline \hline

\multicolumn{3}{|c|}{\cellcolor{gray!25} \textbf{Mappings}} \\\hline \hline
       \multicolumn{3}{|c|}{ VehicleData List (address$\rightarrow$ \textit{VehicleInfo})} \\ \hline

      \multicolumn{3}{|c|}{UAVData List (address$\rightarrow$ \textit{UAVInfo})} \\ \hline

 \multicolumn{3}{|c|}{UAVs in Zone List (ZoneID $\rightarrow$ \textit{address[]})} \\ \hline
\multicolumn{3}{|c|}{UAV Proposal List (UAV address$\rightarrow$ \textit{VehicleSelection[]})} \\\hline
\multicolumn{3}{|c|}{UAV Selection List (UAV address$\rightarrow$ \textit{address[]})} \\\hline
 \multicolumn{3}{|c|}{Selected UAVs (ZoneID $\rightarrow$ UAV address [])}  \\\hline
\end{tabular}
\end{table}

Table \ref{tab:functions} shows the implemented functions as part of the smart contract. $registerUAV()$ and $registerVehicle()$ are for UAVs and vehicles to provide their information in the proposed framework where the smart contract initializes their reputation and creates their respective information structure with its timestamp. Meanwhile, $updateVehicleInfo()$ allows a vehicle to update its information while $updateUAVInfo()$ and $updateUAVZone()$ are for UAVs to update their information and zone. The stored information is used for vehicles to determine their selected UAVs and to use the $submitVehSelection()$ function to submit their UAV proposal information. $submitVehSelection()$ function calculates the \textit{QoV} and $QoV/\sqrt{RB}$, discussed in the following section, and creates the \textit{VehicleSelection} structure to store the information in.  Subsequently, $allocateZone()$ function would be used to determine the selected vehicles for UAVs in a given zone based on submitted proposals. $resetListsForZone()$ resets the list of UAVs in a zone and the UAV selection list while $resetUAVSubmission()$ deletes the entries of the proposal and selection lists of a given UAV. \textit{determineMDRLModel()} function determines which pre-trained MDRL model to deploy on the UAVs, depending on attributes such as the number of agents and number of vehicles.
\begin{table}[h]
\centering
\small
\caption{Functions}
\label{tab:functions}
\begin{tabular}{|c|p{1.3in}|p{0.5in}|} 
\hline
\centering \cellcolor{gray!25} \textbf{Function} & \cellcolor{gray!25} \centering \textbf{Parameters} & \cellcolor{gray!25}  \textbf{Return}\\\hline \hline
$registerUAV()$&\centering UAVInfo parameters  & - \\ \hline
$registerVehicle()$&\centering VehicleInfo parameters  & - \\ \hline \hline
$updateVehicleInfo()$&\centering VehicleInfo parameters & -  \\ \hline
$updateUAVInfo()$&\centering UAVInfo parameters & -  \\ \hline
$updateUAVZone()$&\centering UAV address and Zone & -  \\ \hline \hline
$submitVehSelection()$ & \centering UAV address, QoV, RB & - \\ \hline
$allocateZone()$ &\centering Zone ID & selections \\ \hline \hline
$resetListsForZone()$ & \centering Zone ID & - \\ \hline
$resetUAVSubmission()$ & \centering UAV address & - \\ \hline \hline

$determineMDRLModel()$ &\centering Info & Model ID  \\ \hline
\end{tabular}
\end{table}

\subsection{UAV-assisted IoV Construction}
\label{UAV-assisted IoV Construction}

\begin{table}[!h]
        \centering
        \small
        \caption{Symbols and Notations}
        \begin{tabular}{|l|l|}
            \hline 
            \cellcolor{gray!25}Symbol & \cellcolor{gray!25}Description \\
            \hline \hline
            $v$ & A vehicle \\ 
            $u$ & A UAV \\ 
            $U$ & List of Available UAVs\\ 
            $V$ & List of Available vehicles\\ 
            $RB_v$ & Requested Bandwidth by $v$ \\
            $AB_u$ & Available Bandwidth by $u$ \\
            $Rep_{u/v}$ & Reputation of \textit{u} or \textit{v} \\
            $PayPerMbps_{v}$ & Average payment per Mbps \\
            $BL_u$ & Battery Level of \textit{u} \\ 
            $Distance_{uv}$ & The Euclidean Distance between \textit{u} and \textit{v} \\ 

            $Zone(v),Zone(u)$ & Zone of \textit{u} or \textit{v} \\ 
            \hline
        \end{tabular}
        \label{tab:notation}
    \end{table}
In this work, QoU and QoV metrics are proposed to identify high quality vehicles and UAVs during the selection process to form high quality IoV. The selection process assumes the allocation runs at an instance of time where the positions of vehicles and UAVs are constant despite their mobility for simplicity. The proposed \textit{QoU} metric is computed locally at vehicles to evaluate relay UAVs in their zone. It combines the UAVs' available bandwidth, energy level, reputation, and distance between the UAV and the vehicle. Vehicles acquire UAV  metrics by accessing the \textit{UAV List} and \textit{UAVData List} variables in the proposed smart contract. As these metrics are of different range, each of them is normalized given its maximum value to be in the range ([0-1]). Then, \textit{QoU} is calculated as the weighted sum of the individual metrics. The \textit{QoU} of UAV \textit{u} for vehicle \textit{v}, $QoU_{uv}$, is calculated based on Eq. \ref{eq:qou}. 
\begin{align}
\label{eq:qou}
QoU_{uv} = 100 \times \left( (w_1\times \frac{AB_u}{\text{Max AB}}) + (w_2\times \frac{BL_u}{\text{Max BL}}) + (w_3\times \frac{Rep_u}{\text{Max Rep}})+ (w_4\times (1-\frac{Distance_{uv}}{\text{Max distance}}))\right)
\end{align}
where the sum of $w_1$, $w
_2$, $w_3$, and $w_4$ is equal to 1. The QoU is formulated to be higher for UAVs with higher available bandwidth to increase the probability of selection, higher energy to reduce UAV re-selection, higher reputation as it implies their reliability, and lower distance to reduce the UAV travel time.   

The proposed \textit{QoV} metric is used by UAVs to evaluate vehicles before selecting them as relay nodes for routing. It combines vehicles' requested bandwidth, reputation, payment per Mbps, and distance from the UAV. UAVs acquire vehicle metrics by accessing the \textit{Vehicle List} and \textit{VehicleData List} variables in the proposed smart contract. Similar to the metrics used for QoU, the metrics of the QoV are normalized given the maximum values and the weighted sum is used to aggregate the metrics. The QoV of vehicle \textit{v} for UAV \textit{u}, $QoV_{uv}$,is calculated based on Eq. \ref{eq:qov}.
\begin{align}
\label{eq:qov}
QoV_{uv} = 100 \times \left(( w_5 \times \frac{RB_v}{\text{Max RB}}) + (w_6 \times \frac{PayPerMbps_v}{\text{Max PayPerMbps}}) + (w_7 \times \frac{Rep_v}{\text{Max Rep}})+ (w_8 \times (1- \frac{ \text{Distance}_{uv}}{\text{Max distance}}))\right)
\end{align}
where the sum of $w_5$, $w_6$, $w_7$, and $w_8$ is equal to 1. The QoV is formulated to be higher for vehicles of higher requested bandwidth, higher payment per Mbps, higher reputation, and of lower distance between vehicles and UAVs. It is worth noting that the weights selected can vary based on whether the IoV is constructed for general use where equal weights can be used or tailored to specific applications where higher weight can be set for metrics of higher importance.

The UAV selection runs in two-stages: 1) \textit{Vehicles Selection} where vehicles select UAVs that they want to route for them, 2) \textit{UAV selection} where UAVs confirm the vehicles they will route for within a specific zone. Algorithm \ref{alg:vsm} reflects the first stage, \textit{vehicle selection}, executed at each vehicle in a zone. The vehicle starts by retrieving the Information of UAVs in its zone \textit{ZoneUAVInfo} through the \textit{UAVs in Zone List} and \textit{UAVData List} mappings. Next, a vehicle goes through the list of UAVs and calculates the \textit{QoU} for each of them based on the acquired info. The UAV with maximum QoU is identified along with its Ethereum address. The address would be then used along with the requested bandwidth value to call the \textit{submitVehSelection()} function in the smart contract.

Algorithm \ref{alg:usm} presents the UAV selection mechanism executed at the \textit{allocateZone()} function. The algorithm runs periodically for zones in the \textit{Zones} List. The algorithm iterates through each UAV $u$ in the zone in the order of their occurrence in the List. First, the $UAVProposalList_u$ is sorted based on the $QoV/\sqrt{RB_v}$ metric. Next, the sorted list is iterated through where the requested bandwidth of a vehicle is checked against the available bandwidth. If it is lower or equal to the available bandwidth, the address of $v$ is added to the $UAVSelectionList_u$ and the available bandwidth is updated. 


Regarding computational complexity, up to $|V|\times |U|$ proposals can be submitted, with each proposal being executed in $O(1)$ time through the mapping structure in the smart contract. Consequently, the run-time of the selection algorithm is $O(|V|\times|U|)$.

\begin{minipage}{0.46\textwidth}
\begin{algorithm}[H]
	\caption{Vehicle Proposal Selection}
	\label{alg:vsm}
 \small
	\begin{algorithmic}[1]
           \STATEx \textbf{input: } \textit{Z(v)}: Zone of vehicle\textit{ v}, \textit{ZoneUAVInfo}: Information of the UAVs in  \textit{Z(v)}
            \STATEx \textbf{output:} $max UAV address$
            \STATE retrieve $ZoneUAVInfo$ using \textit{Z(v)}, \textit{UAVs in Zone List}, and \textit{UAVData List} from the smart contract
            \STATE $max QoU = 0$ 
            \STATE $max UAV address = \phi$ 
            \FOR{ $u \in ZoneUAVInfo$}
            \STATE Calculate $QoU_{uv}$ using Eq. \ref{eq:qou}
            \IF{$QoU_{uv} > max QoU$}
            \STATE $max QoU=QoU_{uv}$
            \STATE $max UAV address =$ address of \textit{u}
            \ENDIF
       \ENDFOR
	\end{algorithmic}
\end{algorithm}
\end{minipage}
\hfill
\begin{minipage}{0.46\textwidth}
\begin{algorithm}[H]
	\caption{UAV Selection Mechanism}
	\label{alg:usm}
 \small

	\begin{algorithmic}[1]
             \STATEx \textbf{input:} $UAV Proposal List_u$: The list of proposing vehicles for $u \in ZoneUAV(z)$ 
            \STATEx \textbf{output:} $UAV Selection List_u$: The list of selected vehicles for $u \in ZoneUAV(z)$  
                \FOR{$u \in Z$}
                \STATE Sort $UAV Selection List$ based on $QoV_{uv}/\sqrt{RB_v}$ in descending order
                    \FOR{$v \in UAV Selection List$}
                    \IF{$RB_v <= AB_u$ }
                        \STATE push address of $u$ to $UAV Selection List_u$ 
                        \STATE $AB_u= AB_u - RB_v $  
                    \ENDIF            
                \ENDFOR
        \ENDFOR
	\end{algorithmic}
 \end{algorithm}
\end{minipage}

\subsection{MDRL Approach for UAV Coordination}

The task execution process in the proposed system is enabled through Multi-Agent Deep Reinforcement Learning (MDRL). Given the selection of UAVs to vehicles done in the previous stage, a team of UAVs in a given zone is tasked with coordinating to provide connectivity to the available vehicles in the area. To maintain connectivity and good coverage, a UAV is required to move in the area intelligently according to the locations of the vehicles, as well as the locations of other UAVs. Each UAV is tasked with maintaining coverage to its selected vehicles, while enhancing or maintaining connectivity with the other UAVs. This task is challenging due to the dynamic nature of the environment, where vehicles are continuously moving. This calls for collaboration and coordination between the UAVs, as well as intelligent decision making based on the locations and distribution of vehicles. It is worth mentioning that each zone has its own coordination task among its selected UAVs, determined by the relay selection mechanism. If a vehicle leaves a zone, the proposed mechanism reassigns the vehicle to a UAV in a different zone, which has its own coordination task. In this work, we focus on the coordination between UAVs in a given zone assuming vehicles only move within the zone without leaving it. Given the problem's nature of sequential decision-making based on the collected observations, MDRL comes as an efficient method to obtain intelligent agents. In MDRL, agents develop their intelligence by learning decision-making policies based on their experiences within the environment, aiming to maximize the numerical reward they receive. This section presents the MDRL formulation of the problem and the modeling of the MDRL solutions in terms of observations design, reward engineering, and optimization method.

\hfill
\subsubsection{MDRL Formulation and Policy Optimization}
\label{MDRL Formulation}
\hfill

In the context of MDRL, Markov Games (MGs) are generally used to extend Markov Decision Processes (MDPs) into multi-agent settings \cite{gronauer2021multi, alagha2023blockchain, alagha4753209blockchain}. In the UAV-based coverage and connectivity problem, the different environment states are expressed by the distribution of agents (UAVs) and vehicles. A MG is defined by a set of $\mathcal{S}$ finite states, finite action sets $\mathcal{A}_1 , \mathcal{A}_2, ... , \mathcal{A}_N$ for each of the $\mathcal{N}$ agents, finite observation sets $\mathcal{O}_1 , \mathcal{O}_2 , ... , \mathcal{O}_N$, a state transition function $\mathcal{P}(s',\textbf{o}$ $|$ $s,\textbf{a})$ that computes the probability of ending up in state $s'$ with observation $\textbf{o}$ after executing action $\textbf{a}$ in state $s$, a reward function $\mathcal{R} : \mathcal{S} \times \mathcal{A} \rightarrow \varmathbb{R}$, and a discount factor $\gamma \in [0,1]$. Here, $\textbf{a} = (a_1,...,a_N)$ and $\textbf{o} = (o_1,...,o_N)$ denote joint actions and observations from the $\mathcal{N}$ agents at a given instant. In MDRL settings, the UAV-based coverage and connectivity problem unfolds over discrete steps. At each step, agent $i$ uses its policy $\pi_i : \mathcal{O}_i\times\mathcal{A}_i \rightarrow [0,1]$ to translate an observation $o_i \in \mathcal{O}_i$ into an action $a_i \in \mathcal{A}_i$, and receives a reward $r_i$.

Using the collected experiences, Proximal Policy Optimization (PPO) \cite{schulman2017proximal} is used in this work to improve the decision-making policies of the agents. PPO is a Policy Gradient (PG) method that has two components, an actor and a critic. The actor (policy) network takes the current observations as input and produces a probability distribution over the possible actions. The critic (value function ) predicts the future rewards, which is used in updating the actor network. The objective is to optimize the actor policy $\pi_\theta$, parametrized by $\theta$, to maximize the sum of rewards in an episode. PPO strikes a balance between simplicity and performance by using a clipped surrogate objective that ensures stable and efficient policy updates.

Algorithm \ref{alg:MDRL} describes the learning under PPO for MDRL following the Centralized-Learning \& Decentralized-Execution (CLDE) method \cite{baker2019emergent, alagha2022target, alagha2023multi, alagha4872731adaptive}. The learning takes place over episodes of the problem, where the environment is reset to a different initial state at the beginning of each episode. During episode step $j$, each agent $k$ uses their observations ($o_k^j$) in a copy of the actor network to generate a probability distribution over possible actions ($P_k$), from which an action ($a_k^j$) is sampled. The agents then perform their joint actions ($\textit{\textbf{a}}^j$) in the environment and collect and store the subsequent observations ($\textit{\textbf{o}}^{j+1}$), reward value ($r^j$), and a termination flag ($f^j$) indicating if the episode has ended. Upon the termination of an episode ($f^j$ equals 1), a new episode is initiated with a reset. An episode terminates if the task is completed successfully or if the maximum episode length is reached. When the number of training timesteps reaches the horizon ($H$), the actor and critic networks are updated using the gathered experiences. This process continues until the predetermined total number of training timesteps is achieved.

\begin{minipage}{0.46\textwidth}

\end{minipage}
\hfill
\begin{minipage}{0.46\textwidth}
\centering
\begin{algorithm}[H]
	\caption{PPO- and CLDE-based MDRL}
	\label{alg:MDRL}
 \small
	\begin{algorithmic}[1]
           \STATEx \textbf{input: } Environment and policy networks
            \FOR{$i$ = 0,1,...,MaxSteps} 
            \STATE $\textbf{o}^0$ = ResetEnvironmnet()
            \FOR {$j$ = 0,1,\dots,MaxEpisodeLength}
            \FOR {$k$ = 1,2,\dots,NumberOfAgents}
            \STATE $P_k$ = Actor($o^j_k$)
            \STATE $a^j_k$ = Sample($P_k$)
            \ENDFOR
            \STATE $\textbf{a}^j$ = [$a^j_1, a^j_2, ...$]
            \STATE $\textbf{o}^{j+1}, r^j, f^j$ = Step($\textbf{a}^j$) \color{blue}{\#execute actions}\color{black}
            \STATE Store\_Experiences()
            \IF{$i$ \% $H$ == 0}
            \STATE Update\_PPO()
            \ENDIF
            \IF{$f^j == 1$}
            \STATE break
            \ENDIF
            \ENDFOR
            \ENDFOR
	\end{algorithmic}
\end{algorithm}
\end{minipage}
\hfill
\vspace{1em}

\begin{figure*}[h]
    \centering
    \includegraphics[width=0.8\textwidth]{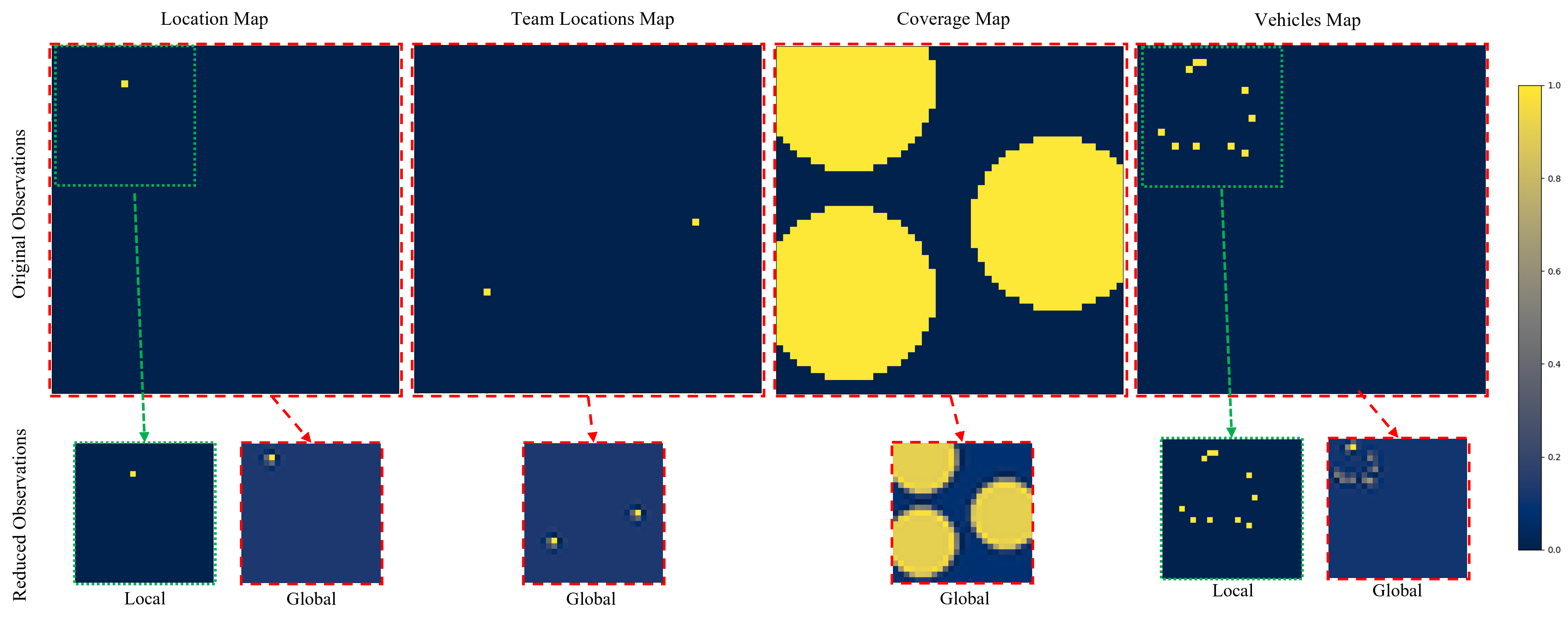}
    \caption{The original observations acquired by a UAV, in a team of 3 UAVs, and their reduced counterparts (local in green and global in red).
    \label{FigObservations}}
\end{figure*}

The decision-making for a UAV at a certain timestep requires knowledge of its own location relative to the zone, the locations and coverage provided by other UAVs, as well as the locations of the vehicles. Due to the spatial nature of the problem, the observations collected by a UAV agent are modeled as 2D grid maps of size $h\times w$ capturing different information about the UAVs and the vehicles. In this problem, and at a given timestep $t$, the observations of UAV $i$ are contained in a stack of four 2D maps as shown in Fig. \ref{FigObservations}. The \textit{Location} map highlights UAV $i$'s location in the zone, the \textit{Team Locations} map highlights the distribution of the other $N-1$ UAVs in the zone, the \textit{Coverage} map shows the areas covered by the team of UAVs, and the \textit{Vehicles} map shows the locations of the vehicles assigned to UAV $i$.

In these observations, the two locations maps help the UAV correlate its own location with respect to the zone and the locations of the other UAVs. The coverage map helps in understanding the coverage provided by the team of UAVs, which is essential in making the right decisions to maintain or enhance connectivity between UAVs based on their ranges. The vehicles map helps the UAV make the right decisions to maintain coverage to the selected vehicles. Here, each UAV has its selected vehicles, and hence its own vehicles map. For maximized connectivity, the UAVs should be positioned so that every UAV can communicate with any other UAV, either directly if they are within communication range, or indirectly by relaying messages through other UAVs in the network. Connectivity could be broken if UAVs move out of the communication range such that there are insufficient relay UAVs to maintain links between distant UAVs. On the other hand, maximized coverage entails that all vehicles are within the range of their selected UAV.

To better optimize the MDRL process, the aforementioned observations undergo pre-processing steps before being fed to the MDRL model. These steps aim to reduce the dimensionality of the observations while maintaining the important information, which results in faster learning and lower computational complexity. Here, the 4 original observations are reduced into 6 smaller observations, as presented in Fig. \ref{FigObservations}. Given the original grid maps of size $h\times w$, the new observations have a $n \times n$ dimension, where $h, w > n > 1$ and $n$ is odd. In this work, $n$ is considered a tunable hyperparameter, where smaller values entail lower computational complexity and higher information loss, when compared to higher values. To maintain essential information when reducing the observations, the reduced observations fall under one of two types: Local and Global. Local observations help the UAV make decisions locally based on the surrounding observations, while global observations help make decisions based on the collective status of the environment. At timestep $t$, local observations are acquired by capturing a $n \times n$ window centered at the UAV's position. In cases where the UAV is near the boundaries of the zone, the center of the window is shifted accordingly to meet the boundaries (as shown in Fig. \ref{FigObservations}). In terms of local observations, the \textit{Local Location} map shows the location of the UAV with respect to the window, while the \textit{Local Vehicles} map shows the distribution of vehicles within the window. The local observations should help the UAV maintain the coverage with the selected vehicles. Ideally, the UAV should learn to maintain a central position with respect to the vehicles, to keep a strong connection and anticipate the movement of vehicles outside its range. On the other hand, the global reduced observations are acquired by applying down-sampling to reduce the original observation into a dimensionality of $n \times n$. In the global observations, the \textit{Global Location}, \textit{Global Team Locations}, \textit{Global Coverage}, and \textit{Global Vehicles} maps are acquired by downsampling the original corresponding maps. The global observations mainly help a UAV cooperate with other UAVs. Here, the spatial trends in the observations are maintained despite the slight loss of information due to downsampling. By observing the coverage, a UAV should learn to get closer to other UAVs to enhance connectivity. At the same time, it is essential to maintain a global observation of the selected vehicles, such that the UAV does not violate its local coverage while attempting to enhance the connectivity. In summary, a UAV's dilemma in this work is to maintain good coverage to its selected vehicles, while attempting to enhance the connectivity with other UAVs. These two objectives could sometimes collide, depending on the locations of UAVs and vehicles, as well as the communication ranges of UAVs, which makes the MDRL problem complex.

\hfill
\subsubsection{Action Space}
\hfill

In the proposed work, a UAV makes decisions in a discrete fashion. At each timestep, the UAV chooses to either move in a specific direction or remain stationary. For simplicity, it is assumed that the UAVs operate in a 2D plane. The action space is divided into $K$ possible directions \{1, 2, ..., $k_i$, ..., $K$\}, where the movement direction (or angle) is computed as: 

\begin{equation}
\theta = 2\pi \frac{k_i}{K}
\end{equation} 
In this work, the UAV has constant speed and 9 possible actions, 8 of which are cardinal and ordinal directions ($K=8)$ and one action for remaining stationary.

\hfill
\subsubsection{Policy Architecture}
\hfill

\begin{figure}[h]
    \centering
    \includegraphics[width=0.8\columnwidth]{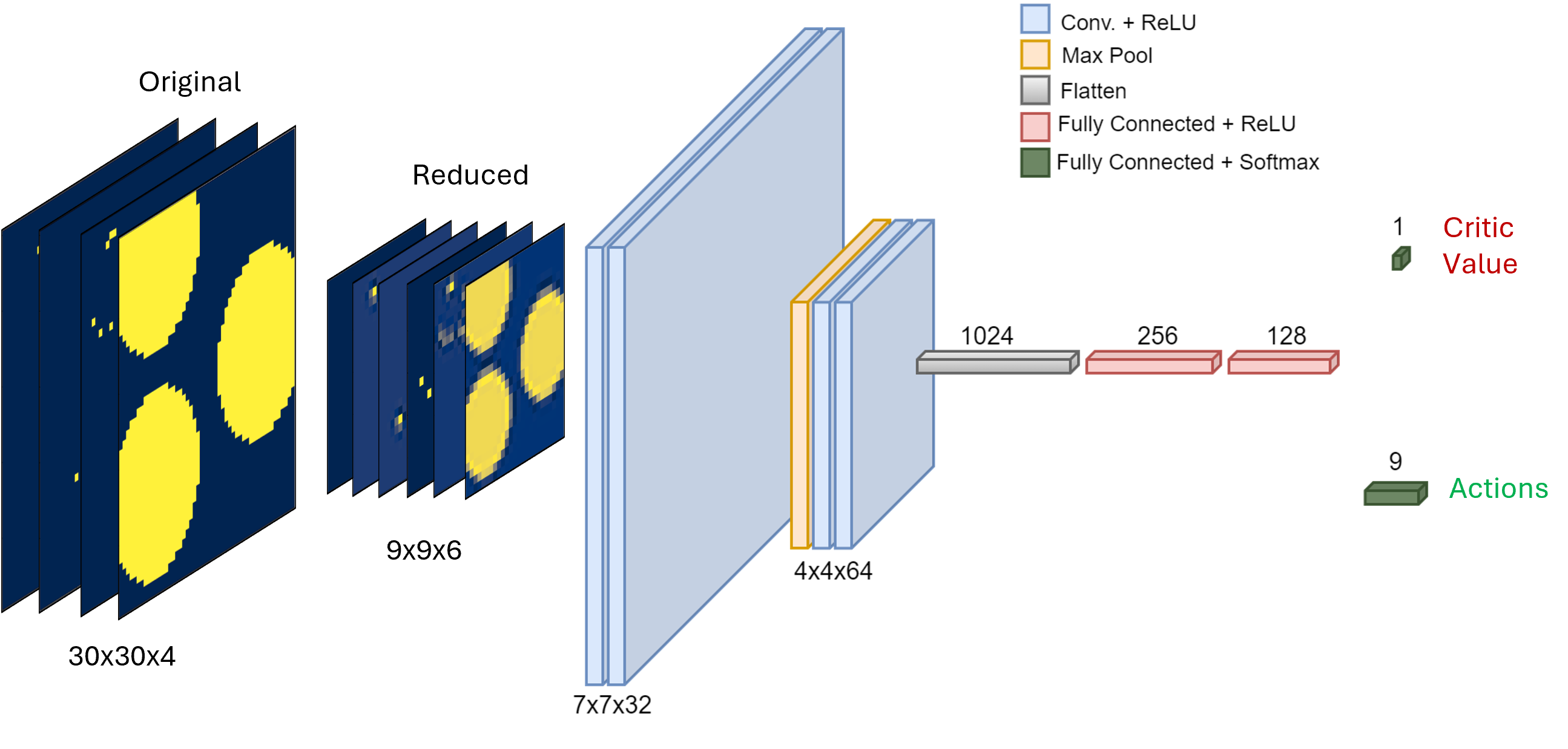}
    \caption{The architecture used for the actor and critic networks. While both networks have the same architecture configuration, they are two separate networks with two different outputs. \label{FigActCrit}}
\end{figure}

The intelligent decision making in the proposed system is done through a policy that is defined as a Convolutional Neural Network (CNN), which is ideal to capture spatial correlations in the collected observations. Fig. \ref{FigActCrit} shows the CNN architecture used for the actor and critic networks, which is based on the LeNet-5 architecture \cite{lecun2015lenet}. The actor network translates the UAV's reduced observations into a probability distribution, through the softmax function, over the 9 possible actions. The critic network takes the reduced observations and produces an estimate to the value function to update the networks. In the proposed CLDE training method, each UAV agent acts based on a copy of the actor, and the acquired experiences and rewards are then used to centrally update a common policy, before being distributed to the agents again. After the training is completed, the agents are deployed on UAVs, where each UAV gets a copy of the policy network and operates independently based on its own observations. It is worth mentioning that the input maps to the actor and critic are typically normalized for optimal performance.

\hfill
\subsubsection{Reward Function}
\hfill

The reward function is essential in guiding the training and ensuring fast convergence. To achieve this, a \textit{shaped} and \textit{joint} reward function is used. A shaped function ensures that the agents get feedback more frequently throughout the episode. This is unlike a sparse reward function that only gives feedback very few times throughout the episode, usually only when the episode is completed successfully. A joint reward function is one where all the agents receive the same reward based on the collective behavior. Here, following a set of actions by the agents in a timestep, all the agents receive the same reward based on the new state of the environment. This pushes the agents to take actions that benefit the team, which is essential in collaborative tasks.

At a given timestep $t$, after the agents have executed the joint action $\textbf{a}^t$, the environment returns a reward $R_t$ based on the following function:

\begin{equation}
    R_t = (coverage - 1) + (connectivity - 1)
    \label{eq:rewardequation}
\end{equation}

In this function, \textit{coverage} is computed as the portion of vehicles, out of all vehicles, which are covered by their corresponding UAVs, which results in a value between 0 and 1. A vehicle is considered to be covered if it lies within the communication range of its allocated UAV. The \textit{connectivity} metric has a binary value (either 0 or 1), where 1 indicates that all the UAVs in the network are linked, and 0 indicates otherwise. The UAVs are considered to be linked if there is a communication path connecting each two UAVs in the network, either directly (i.e. within each other's communication range) or indirectly (through intermediate UAVs that are within each other's communication range). In Eq. \ref{eq:rewardequation}, each attribute has a penalty of (-1) value, which generally helps achieving the desired outcome faster while training. Hence, the maximum value in a time-step is 0, indicating full connectivity and coverage, while the minimum is -2, indicating no connectivity and no coverage.


\subsection{Process Sequence Diagram}

\begin{figure}[h] 
\centering
\includegraphics[width=0.5\linewidth]{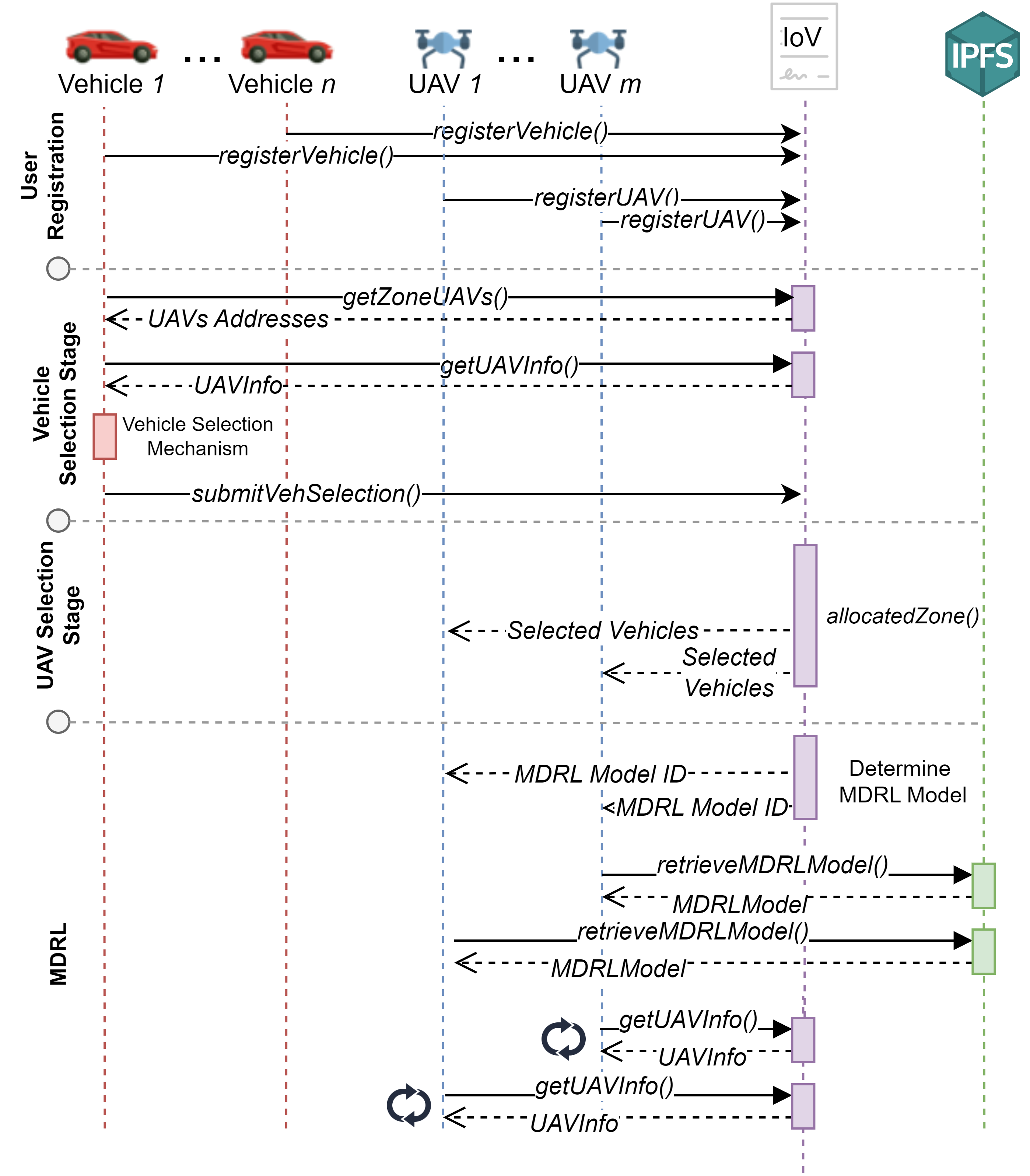}
\caption{The overall process diagram}
\label{fig:process}
\end{figure}

Figure \ref{fig:process} shows the interactions between the different components in the proposed framework. It presents the interactions between vehicles, UAVs, the proposed smart contract and the IPFS system. The interaction encapsulates the following stages:

\begin{itemize}
    \item \textbf{User Registration.} Vehicles and UAVs register to \textit{IoV} by providing the required information to the \textit{registerVehicle()} and \textit{registerUAV()} functions respectively. The respective function encapsulates the information in a \textit{vehicle} or \textit{UAV} object and appends the object to the respective list \textit{Vehicle List}/ \textit{UAV List}. Each vehicle/UAV is assumed to have a single Ethereum address linked to its account for the reputation. 
    \item \textbf{Vehicle Selection Stage.} Vehicles retrieve the UAVs in their zone by invoking the \textit{getZoneUAV()} function, which returns the addresses of UAVs as a list. Then, the vehicle can get the detailed information about the UAVs by sequentially invoking the \textit{getUAVInfo()} function with the address of each UAV and retrieving the UAVInfo. Once the UAV information is retrieved, the vehicle runs the selection mechanism proposed in Algorithm \ref{alg:vsm} to determine the UAV it selects. The selection is then submitted through the \textit{submitVehSelection()} function.
    \item \textbf{UAV Selection Stage.} The \textit{allocateZone()} function is executed to select UAVs for each zone based on the submitted proposals by vehicles according to Algorithm \ref{alg:usm}. Upon the completion of the selection, UAVs are notified about vehicles they are expected to route for. 
    \item \textbf{Coordination.} Upon selection of vehicles to UAVs, the coordination stage is initiated. \textit{IoV} determines the MDRL model to dispatch based on the characteristics of the UAV and returns an identifier to retrieve it from IPFS. Once the MDRL model is retrieved, the UAV can follow the actions suggested by the model based on its observations to optimize its location, while continuously retrieving the information of other UAVs from \textit{IoV}.
\end{itemize}

During the coordination phase, certain practical challenges, such as UAV battery drainage, can impact the sustainability of the system. While the proposed framework focuses on optimizing coverage and connectivity through efficient coordination, it is essential to acknowledge that UAVs cannot remain active indefinitely due to energy constraints. To address this, one simple solution is to replace depleted (or malfunctioning) UAVs with ones from the available pool through UAV selection, following the process discussed in Section \ref{UAV-assisted IoV Construction}. Alternatively,  the proposed method can be complemented through existing solutions that address energy management challenges, such as deploying charging stations to enable UAVs to recharge during operations \cite{fan2022deep, liu2020energy}.

\section{Simulation and Evaluation}
The evaluation conducted for the proposed framework and MDRL method is divided into three main components. First, the performance of the UAV selection mechanism is evaluated at an instance of time. Second, the performance of the MDRL model is analyzed and compared with existing benchmarks. Third, the cost analysis and scalability of the proposed framework and incorporated mechanisms are discussed to verify its cost efficiency and scalability. 

\subsection{Simulation Setup}

Table \ref{tab:setup} summarizes the used evaluation setup for the UAV selection mechanism. Here, the evaluation is conducted on MATLAB 2023b where a dataset of vehicles and UAVs is generated as described in \cite{ABUALOLA2021100290}. A fixed number of vehicles is generated within the simulation area while an increasing percentage of UAVs is generated alongside the vehicles to evaluate the scalability. The attributes of vehicles and UAVs are initialized randomly based on a uniform distribution. The maximum value for each attribute is indicated in the table where the max distance is based on the range of the zone while the other metrics are based on the ranges of the metrics. The weights used in Eq. \ref{eq:qou} and \ref{eq:qov} are set to 0.25 to give equal contribution to the normalized attributes. The simulation area is fixed at 50 km $\times$ 50 km to control the complexity of the environment. By varying the number of UAVs and vehicles in the area, the experiments effectively adjust the density of the agents. This introduces the same level of complexity as fixing the team size and varying the area size, as both scenarios alter the density of agents and influence the dynamics of coverage and connectivity. This allows for a systematic evaluation of the scalability and adaptability of the proposed methods.
 
\begin{table}[!h]
        \centering
        \small
        \caption{Simulation Setup}
        \begin{tabular}{|l|c|}
            \hline 
            \cellcolor{gray!25}Parameter & \cellcolor{gray!25}Value \\
            \hline \hline
            Simulation Area & $50\times 50~km^2$ \\
            Number of Vehicles & 200 \\
            Number of UAVs & [20, $\dots$, 160]  \\ 
            Number of Iterations & 5 \\ \hline \hline
            Latitude \& Longitude & Uniform distribution [0-50] \\ 
            Max distance &  3Km  \\\hline \hline

            $RB_v$ & Uniform distribution [1-4 Mbps] \\ 
            \textit{Max RB} & 4 Mbps \\ \hline \hline

            $Rep_{u/v}$ &  Uniform distribution [1-100]\\  
            \textit{Max Rep} & 100 \\ \hline \hline

            $PayPerMbps_v$&  Uniform distribution [0-7 \$\/Mbps]\\ 
            \textit{Max PayPerMbps} & 7 \\ \hline \hline

            $AB_u$ & Uniform distribution [0-20 Mbps]\\ 
            Max AB & 20 Mbps \\ \hline \hline

            $BL_u$ &  Uniform distribution [1-100]\\ 
            Max BL& 100 \\ \hline \hline
            $w_1,w_2,w_3,w_4,w_5,w_6,w_7,w_8$ & 0.25 \\ \hline
        \end{tabular}
        \label{tab:setup}
    \end{table}

For MDRL, the simulations are performed using Python on an Intel E5-2650 v4 Broadwell workstation, which features 128 GB RAM, an 800 GB SSD, and an NVIDIA P100 Pascal GPU with 16GB of HBM2 memory. For all the experiments, each model is trained for 1 million environment steps using MDRL. At the beginning of each episode, the vehicles are randomly placed in the environment and assumed to be assigned to each UAV based on the selection mechanism. The UAV is placed such that it covers its assigned vehicles. An episode continues for 100 steps, where the aim is to maximize the coverage and connectivity. For each 40,000 training steps, the agents are placed in a testing environment for 4,000 steps where they act greedily (by always taking the most valued action) based on their latest policy. The average performance of the testing steps is then recorded and plotted. The hyperparameters used in the PPO method are summarized in Table \ref{Hyperparameters}, which are based on the original work in \cite{schulman2017proximal}.

\begin{table}[ht]
\caption{Hyperparameters used for PPO training.}
\setlength{\tabcolsep}{0pt}
\begin{center}
\begin{tabular}{|l|P{0.25\columnwidth}|}
\hline
\cellcolor{gray!25}PPO Hyperparameters &\cellcolor{gray!25} Value\\
\hline \hline
Entropy coefficient $c_2$ & $0.01$\\
Learning rate & $3\times 10^{-4}$\\
Discount factor $\lambda$ & $0.95$\\
PPO clipping parameter $\varepsilon$ & $0.2$\\
Number of epochs per update & $20$\\
Discount factor $\gamma$ & $0.99$\\
Timesteps per update (Horizon $H$) & $4000$\\
\hline
\end{tabular}

\end{center}
\label{Hyperparameters}
\end{table}

The same simulation setup was used for both the selection mechanism and the UAV coordination task. However, it is important to note that these two components target distinct aspects of the framework. The selection mechanism operates at a global level, considering the entire area when allocating UAVs to vehicles. This involves dividing the area into zones and assigning UAVs to specific zones based on the selection mechanism. After selection, each team of UAVs acts independently in its allocated zone based on the MDRL model to optimize coverage and connectivity, hence the scope of MDRL is only a single zone. Nonetheless, for both parts, we experiment with varying number of vehicles/UAVs to show the adaptability of the proposed methods to different environmental complexities.


\subsection{Selection Mechanism}


The proposed selection mechanism is evaluated against a modified version of the Nearest Neighbor Matching (NNM) mechanism to compare the resulting matching based on different matching methodologies. NNM utilizes propensity score \cite{dehejia2002propensity} and an acceptable list to create well-matched groups with similar scores  members without considering the preferences of members. To ensure consistency with the proposed mechanisms in terms of matching attributes, the propensity score is replaced with the proposed QoU and QoV metrics. The results demonstrate the performance at an instant of time where the vehicles are assigned UAVs within their zones. The UAVs would then be responsible for maintaining coverage and connectivity throughout the mobility of vehicles. 



\begin{figure}[h!]
     \centering
     \subfloat[\label{fig:vehiclesperUAV} Number of Vehicles Per UAV]{
     \includegraphics[width=0.45\columnwidth]{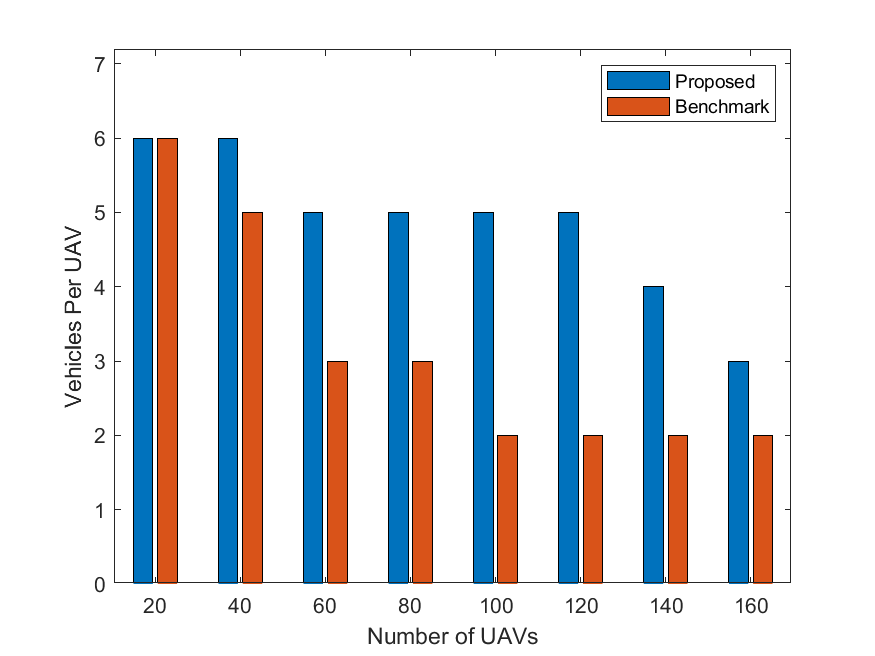}}
     \subfloat[\label{fig:percofuavs} Percentage of UAVs selected]{
     \includegraphics[width=0.45\columnwidth]{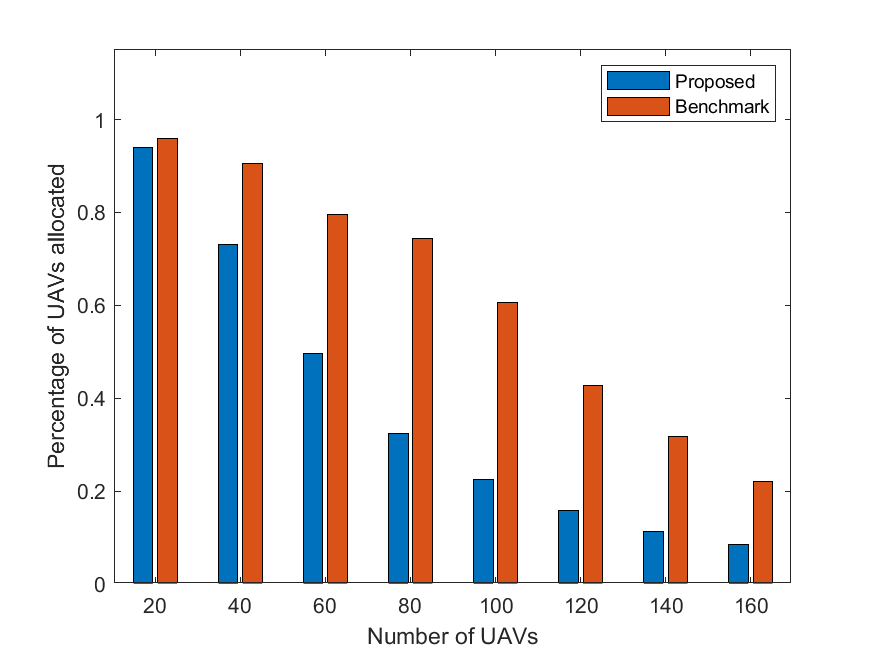}}
     \hfill
        \caption{Number of vehicles accepted per UAV and the percentage of selected UAVs.}
        \label{fig:allocationmech1}
\end{figure}

Figs. \ref{fig:vehiclesperUAV} and \ref{fig:percofuavs} presents the number of vehicles accepted per UAV and the percentage of selected UAVs. It can be seen that for a small number of UAVs ($\leq 20$) both mechanisms have a similar number of vehicles per UAV, but we can notice a slight difference in the percentage of UAVs selected. When the number of UAVs increases ($>20$), the proposed approach still maximizes the number of vehicles selected per UAV while requiring a much smaller percentage of UAVs to be selected compared to the benchmark. The benchmark offers a smaller number of vehicles per UAV as it does only consider the similarity of QoU and QoV values rather than the preference for UAVs and their capacity to server other vehicles, which hence impacts the percentage of UAVs selected when more UAVs are part of the network.

\begin{figure}[h!]
     \centering
     \subfloat[\label{fig:qou} Average QoU]{
     \includegraphics[width=0.45\columnwidth]{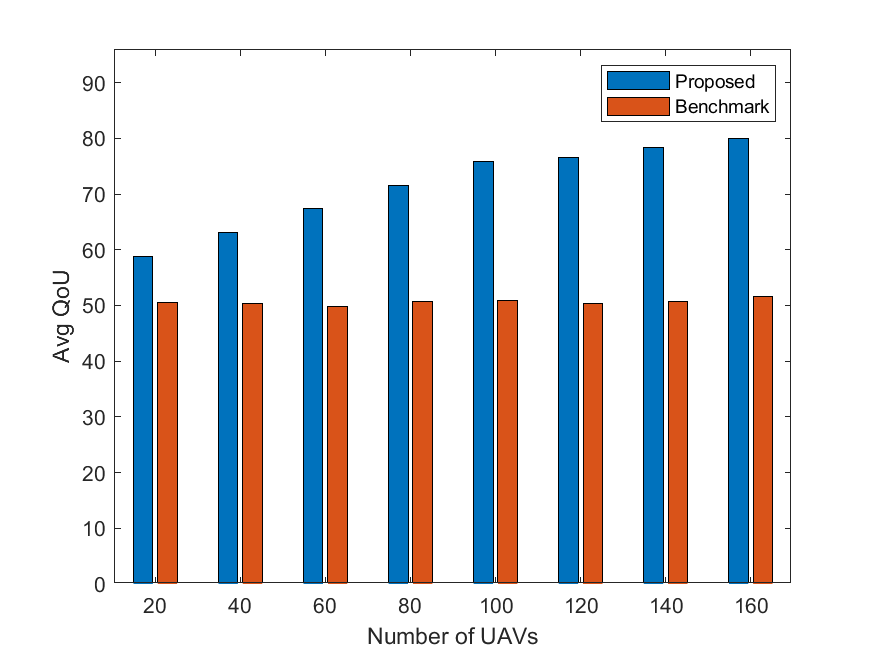}}
     \subfloat[\label{fig:qov} Average QoV]{
     \includegraphics[width=0.45\columnwidth]{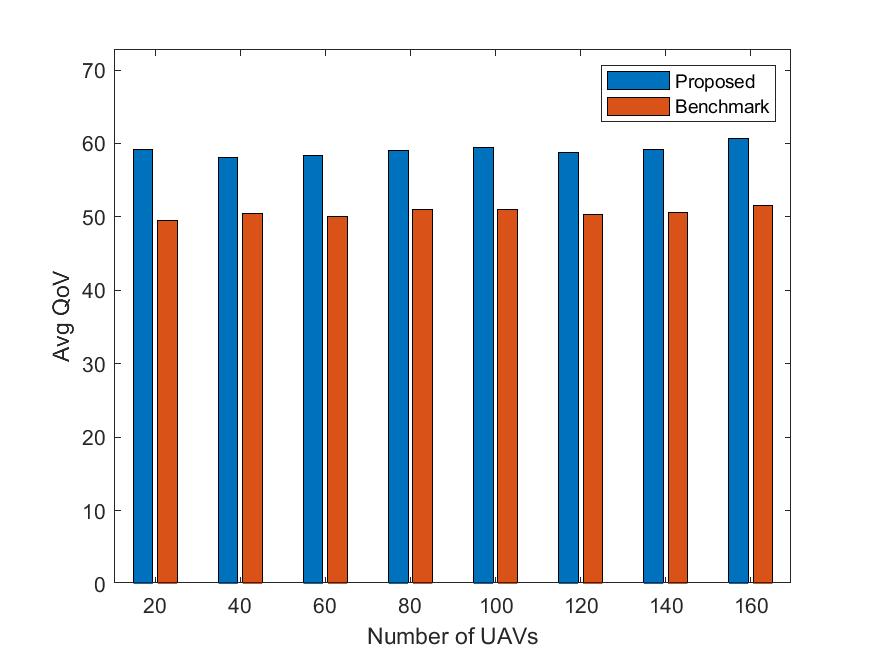}}
     \hfill
        \caption{Average QoU and QoV values achieve based on the selection mechanism.}
        \label{fig:allocationmech2}
\end{figure}

Figs. \ref{fig:qou} and \ref{fig:qov} show the average QoU and QoV values for selected vehicles and UAVs. The proposed mechanism outperforms the benchmark in both QoU of selected UAVs (at least 13\%) and QoV of selected vehicles (at least 10\%). The proposed mechanism demonstrates an increasing average QoU with higher percentage of UAVs available while the benchmark results in a stable performance in terms of the average QoU regardless of the number of UAVs available. Meanwhile, the proposed mechanism demonstrates a consistent increase in the average QoV of selected vehicles compared to the benchmark (an average of 10\%). The higher avg QoU by the proposed mechanism is due to the fact that GSM is optimal to the side initiating the proposals (vehicles) and would maximize the metric they use during selection (QoU). Nevertheless, as it depends on the preferences rather than the similarity, it observes a higher QoV performance compared to its benchmark.

\subsection{MDRL-based Coordination}
\begin{figure}[h!]
     \centering
     \subfloat[\label{2agentrewards}]{
     \includegraphics[width=0.48\columnwidth]{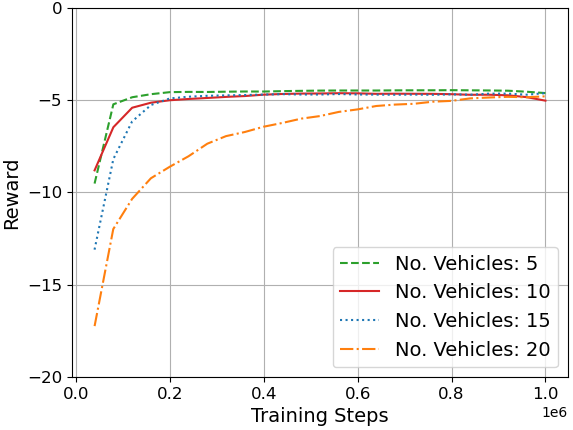}}
     \hfill
     \subfloat[\label{3agentrewards}]{
     \includegraphics[width=0.48\columnwidth]{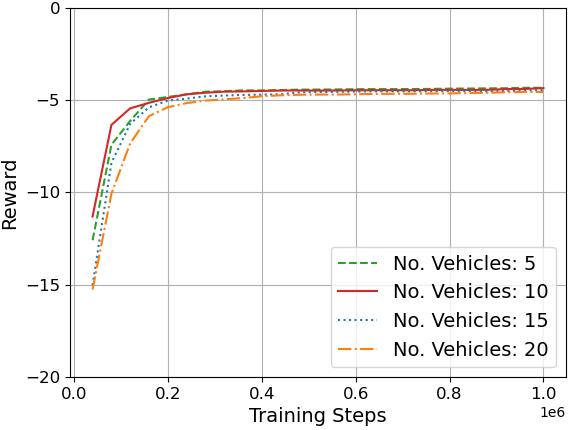}}
     \hfill
     \subfloat[\label{4agentrewards}]{
     \includegraphics[width=0.48\columnwidth]{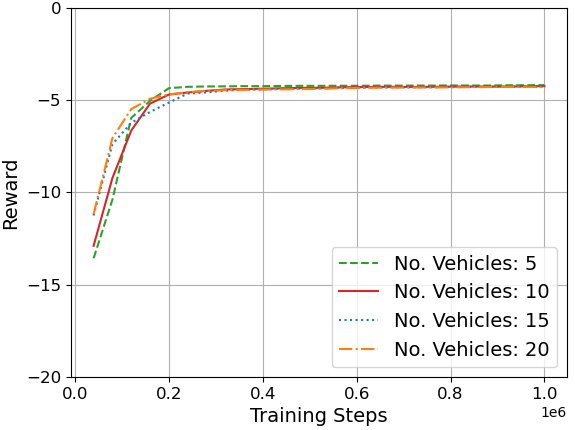}}
     \hfill
     \subfloat[\label{5agentrewards}]{
     \includegraphics[width=0.48\columnwidth]{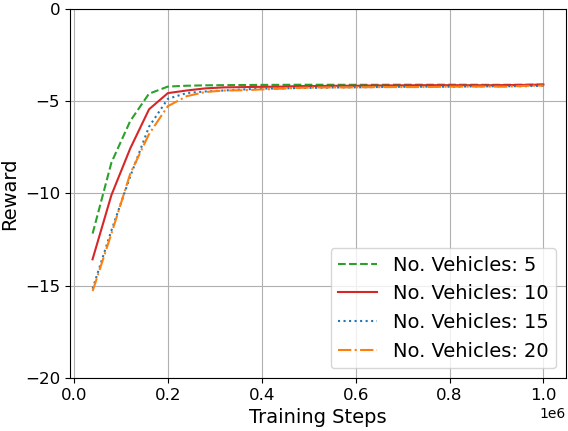}}
     \hfill
        \caption{The average episodic reward throughout the learning for varying numbers of vehicles, for a scenario of (a) two UAVs, (b) three UAVs, (c) four UAVs, (d) and five UAVs.}
        \label{Results:rewards}
\end{figure}

This section validates the performance of the MDRL method under different numbers of agents (UAVs) and different numbers of vehicles. Fig. \ref{Results:rewards} shows the learning curves in terms of episodic rewards for the different scenarios. As evident, regardless of the complexity of the problem, the MDRL method is able to converge in a timely manner. It can be seen that, as the number of vehicles increases, the convergence occurs slightly slower. This is expected since the coverage and connectivity tasks become more complex with increasing number  of vehicles. However, in all situations, the learning converges to a reward value between -4 and -5. Based on the reward function introduced in Eq. \ref{eq:rewardequation}, a negative reward is accumulated if connectivity or coverage are not maximized. Upon inspecting the behaviors of the agents towards the end of the learning, it was evident that the majority of the negative reward is accumulated during the initial steps of the episode. This is mainly because the UAVs are placed randomly, hence they need time to coordinate and achieve full connectivity and coverage. For example, for the case of 5 agents (Fig. \ref{5agentrewards}) where the learning converges to a reward of -4, the agents need 4 timesteps on average to achieve full connectivity and coverage. Full connectivity here means that each UAV has at least one other UAV within its range. The challenge here, which is intelligently tackled by the UAVs, is to achieve connectivity while maintaining their coverage of the assigned mobile vehicles. It is noticed that while the agents attempt to achieve full connectivity, in rare cases, some vehicles might temporarily leave the coverage region of their assigned UAVs. However, this only occurs in the first few steps, and once the UAVs reach full connectivity they are able to maintain full coverage throughout the episode. This reflects the intelligent decision-making developed by the agents through the MDRL process.

\begin{figure}[H]
     \centering
     \subfloat[\label{FigBenchMDRL2}]{
     \includegraphics[width=0.48\columnwidth]{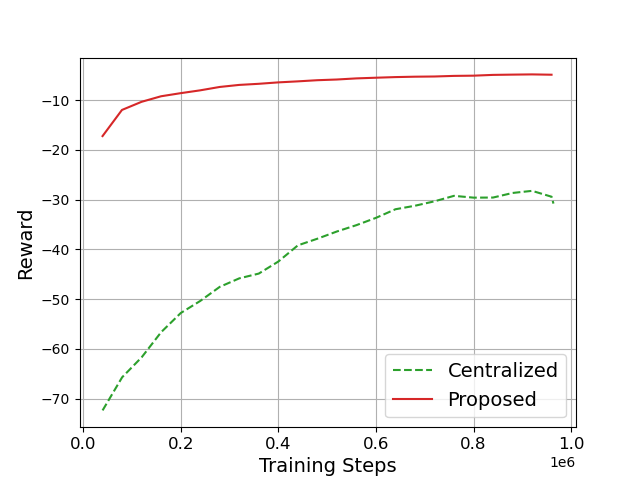}}
     \hfill
     \subfloat[\label{FigBenchMDRL3}]{
     \includegraphics[width=0.48\columnwidth]{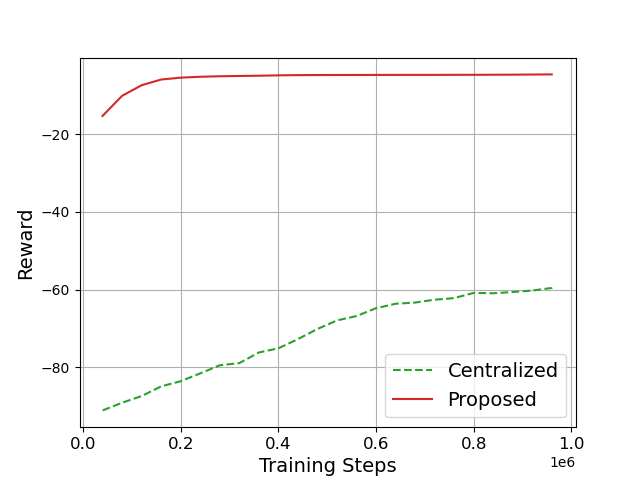}}
     \hfill
        \caption{Comparison between the proposed MDRL and the benchmark for a scenario of 20 vehicles and a team of (a) two UAVs and (b) three UAVs.}
        \label{FigBenchMDRL}
\end{figure}

Figure \ref{FigBenchMDRL} compares the proposed MDRL method with existing literature that uses centralized decision-making  \cite{8960481, 9082162}. While these works use different DRL techniques to translate observations into actions, such as Q-Learning and DDPG, the decision-making is done centrally by a single agent that controls all UAVs. To analyze this behavior, we modified our proposed method to have a centralized nature, where a single agent takes the combined observations of all the UAVs and produces a combination of actions that control their movements. This is used as a benchmark to analyze the performance of centralized methods. Fig. \ref{FigBenchMDRL} shows the performance for two team sizes, namely 2 and 3 UAVs. As can be seen, the benchmark struggles with the learning, especially as the team size grows. This is mainly because of its space complexity, where the action space for a centralized agent grows exponentially with the number of UAVs. For example, for two UAVs each having 9 actions, the agent deals with 81 combinations of these actions, which grows to 729 in the case of 3 agents. This makes the learning significantly harder, as shown in the figure. On the other hand, the agents act in a distributed manner in the proposed MDRL method based on their independent observations, which makes the learning scalable to larger team sizes. This can also be seen when studying the number of trainable parameters in the model as the team size increases. For the proposed method, there are 1.692M trainable parameters regardless of the team size. However, for the centralized method, the number of trainable parameters grows exponentially with more agents, going from 1.692M for 2 UAVs to 9.31M for 5 UAVs, showing its scalability issues. In terms of time complexity, RL algorithms relying on Temporal Difference (TD) error, such as PPO, have a time complexity of $O(d)$, where $d$ is the complexity of the used model (i.e. the neural network) \cite{sutton2018reinforcement}. PPO uses TD in batches and epochs, resulting in a complexity of $O(K \times B\times d)$, where $K$ is the number of epochs and $B$ is the batch size used. Since $K$ and $B$ are hyperparameters, the main influencer on the time complexity here would be the complexity of the policy model $d$. To assess this complexity, we use the FLOPs (Floating Point Operations) metric, which is commonly used in practice to estimate the computational cost of training DL and DRL models, and which can be easily translated into training speed and power consumption depending on the machine used. The FLOPs metric is computed by tracking the total number of floating-point operations by all the layers of a DL model. In this context, the proposed method has approximately 1 MFLOPs ($10^6$) based on the CNN architecture used. This is regardless of the number of agents, which shows its scalability. On the other hand, the FLOPs for the centralized method range from 1 MFLOPs for 2 UAVs all the way to 41 MFLOPs for 5 UAVs, which shows the method's scalability issues.

\begin{figure}[h]
     \centering
     \subfloat[\label{CoverageFig}]{
     \includegraphics[width=0.473\columnwidth]{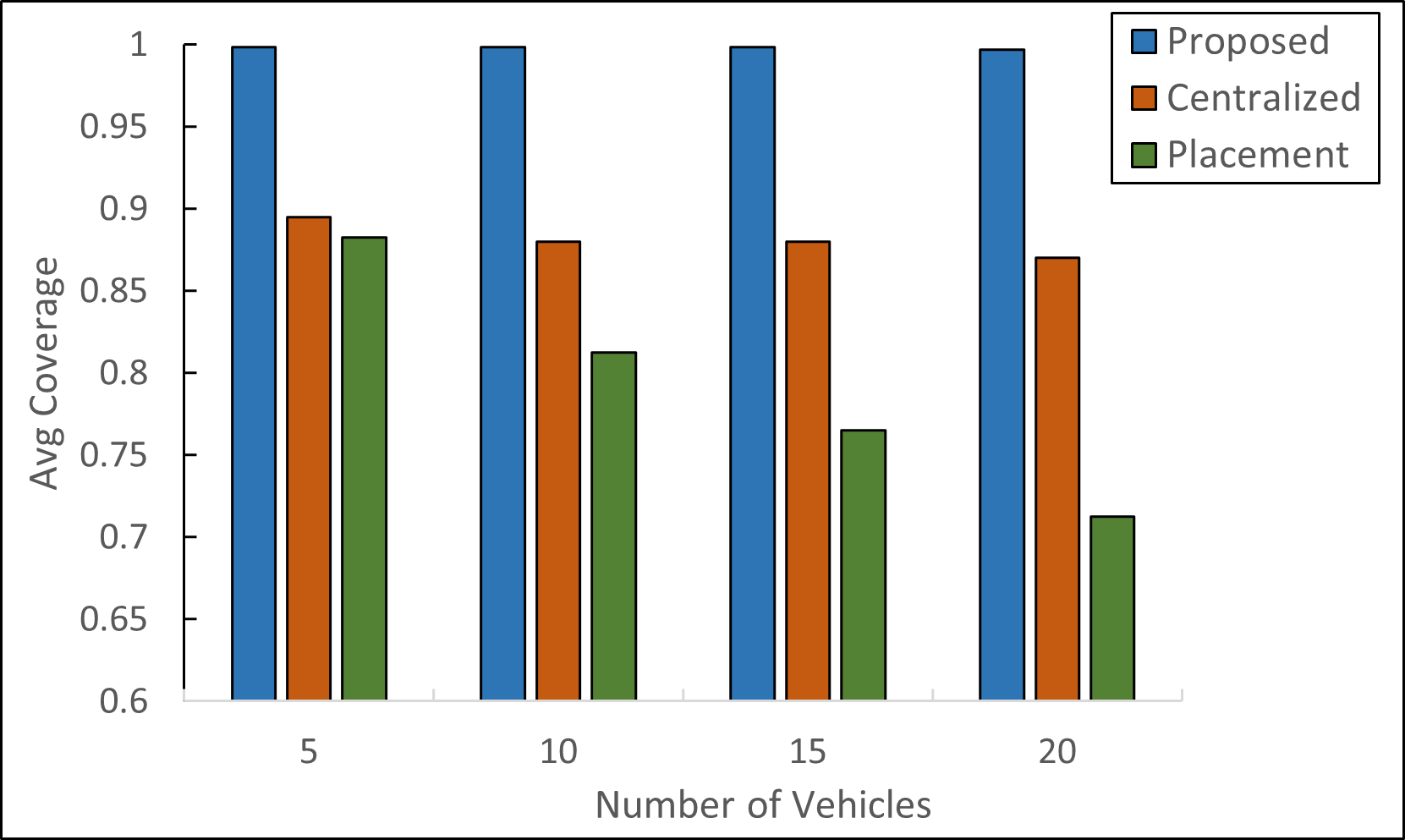}}
     \hfill
     \subfloat[\label{ConnectivityFig}]{
     \includegraphics[width=0.48\columnwidth]{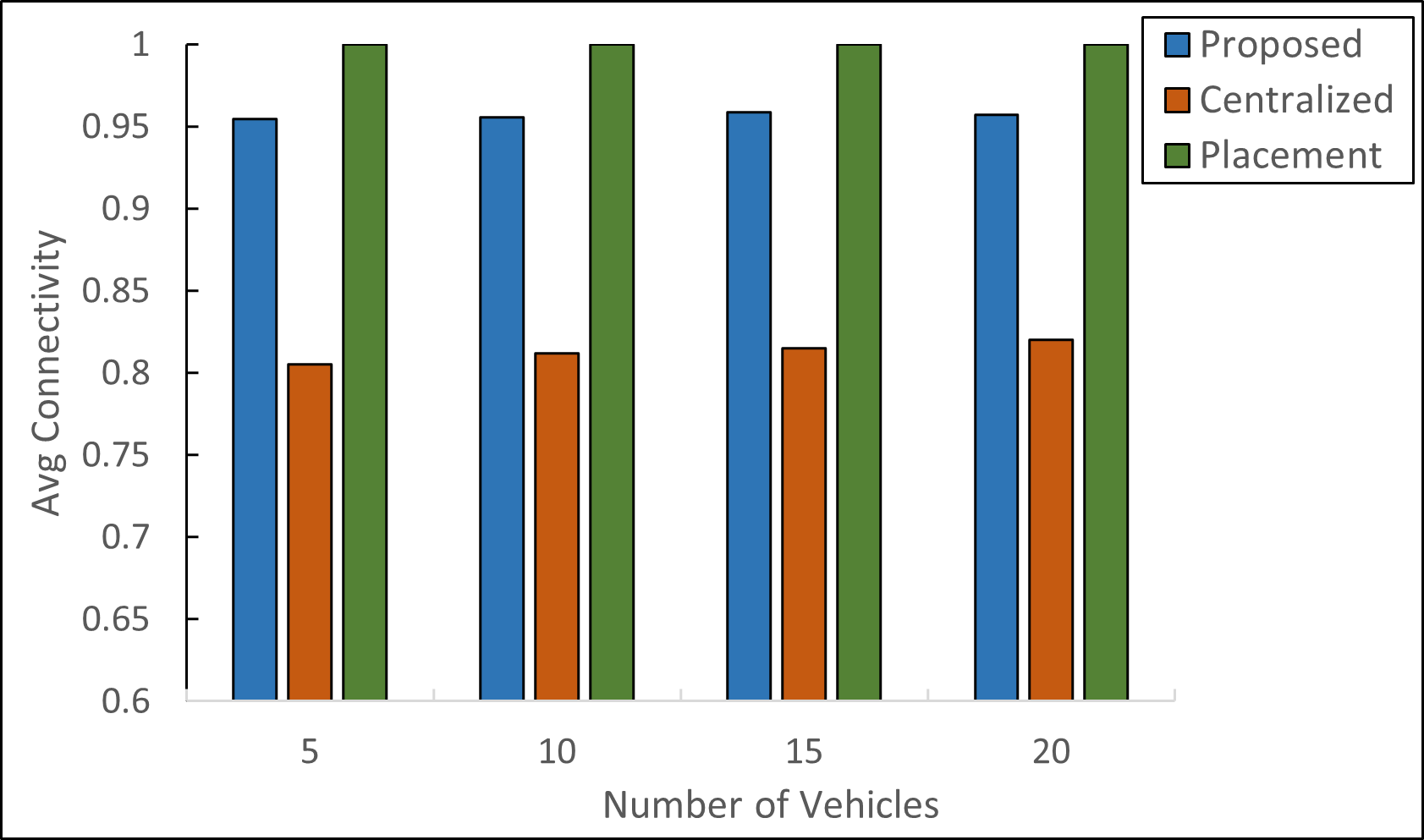}}
     \hfill
        \caption{The coverage and connectivity achieved by different team sizes with different number of vehicles.}
        \label{Results:CovCon}
\end{figure}

Building upon the previous analysis, Fig. \ref{Results:CovCon} further evaluates the systems' performance in terms of connectivity and coverage while considering different number of vehicles. Each data point is the average of 20 experiments where the team size is randomly set between 2 and 5 UAVs. Here, the trained UAV agents are placed in an environment for episodes of 100 timesteps and are tasked with maximizing the coverage and maintaining connectivity. Coverage reflects the portion of vehicles covered per step on average, while connectivity reflects the portion of steps (out of 100) where the UAVs maintained full connectivity with each other (i.e. each UAV has at least one other UAV within its range). The performance of the proposed decentralized MDRL method is compared with two benchmarks, namely centralized DRL and UAV placement. For centralized DRL, the trained agent obtained after 1 million training steps (as per Fig. \ref{FigBenchMDRL} is used to control all the UAVs. For UAV placement, the work in \cite{hadiwardoyo2020three} is extrapolated to consider both coverage and connectivity, where PSO is used to optimize the placement of the UAVs such that connectivity and coverage are ensured. It can be noticed in the figure that the proposed method outperforms centralized DRL in terms of coverage and connectivity, even though both methods were trained using the same amount of experience. This signifies the scalability issue existing in centralized learning when extended to multi-agent setting due to the curse of dimensionality. When analyzing the performance of UAV placement methods, it can be noticed that the coverage significantly deteriorates as the number of vehicles increases, since UAVs cannot adapt to the dynamicity of the vehicles. This is unlike the proposed method where UAVs observe their vehicles and other UAVs and adapt accordingly. In terms of connectivity, placement methods predictably maintain full connectivity since the UAVs maintain a static behavior once positioned. On the other hand, the proposed has a slightly lower connectivity score. This is attributed to the fact that the UAVs, once selected and selected their vehicles, start the coordination tasks without necessarily having connectivity, and hence it is their responsibility to achieve connectivity afterwards. As per the results, it can be noticed that the connectivity score is 0.95 on average, indicating that the UAVs need only 5\% of the episode (5 steps) to achieve connectivity once selected, while also maintaining the coverage of their selected vehicles.

\subsection{Cost Analysis}



This section details the implementation and evaluation of the proposed framework to demonstrate its feasibility. The framework's smart contracts are developed using Solidity 0.8.26 on the Ethereum IDE\footnote{https://remix.ethereum.org/} and deployed in the Remix VM (London) environment. This environment includes multiple test user accounts representing requesters and workers within the framework. The costs associated with deploying the framework and executing its mechanisms are presented in Table \ref{tab:costfunction}. It is assumed that 1 Gas unit costs 6 Gwei for an average confirmation time of 30 seconds\footnote{hthttps://goto.etherscan.com/gastracker}.

\begin{table}[!htb]
	\centering
	\caption{Implementation Cost}
	\label{tab:costfunction}
	\begin{tabular}{ |c|c|c| } \hline 
	 \cellcolor{gray!25}\textbf{Function} & \cellcolor{gray!25}\textbf{Gas} &\cellcolor{gray!25} \textbf{Ether}\\ \hline
	\hline
        Deployment & 2932986  &0.0176 \\\hline
	$registerUAV()$ & 265964    & 0.0016  \\ \hline
        $registerVehicle()$& 161646   & 0.0009 \\ \hline
        $updateVehicleInfo()$&  54002  & 0.0003\\ \hline
        $updateUAVInfo()$&  67287 & 0.0004 \\ \hline
        $updateUAVZone()$&  104492   & 0.0006\\ \hline
        $submitVehSelection()$&  159442  & 0.0009  \\ \hline
        $resetListsForZone()$&  116565 & 0.0006 \\ \hline
	\end{tabular}
\end{table}

The deployment cost is a one-time paid cost to set up the framework smart contract. The smart contract's $registerUAV()$ and $registerVehicle()$ functions are one-time functions called by UAVs and vehicles when registering in the framework. Meanwhile, the cost of update functions, $updateVehicleInfo()$, $updateUAVInfo()$, and $updateUAVZone()$, are called as per required by the members of the IoV with their cost being endured by the calling member. For the selection process, the $submitVehSelection()$ function is paid per vehicle submitting its selection for the UAV it prefers, and the cost is relatively small. The $allocateZone()$ function is paid by all UAVs in the zone being selected vehicles along with the $resetListsForZone()$ function making their amount reasonable for the acquired return.  Overall, the implementation of the smart contract and its functions demonstrates the feasibility of the proposed framework and the cost efficiency of the implemented functions.

\section{Conclusion}
In this article, a comprehensive management system for UAV-assisted IoV networks is proposed, leveraging Blockchain and Multi-Agent Deep Reinforcement Learning (MDRL). The proposed system comprises a Blockchain-based UAV selection mechanisms for relay selection, which matches vehicles to relay UAVs by considering both of their preferences in a transparent and trusted manner. A smart contract is designed to manage the selection based on the proposed Quality-of-Vehicle (QoV) and Quality-of-UAV (QoU) metrics. Once selected, a trained MDRL model is deployed on the UAVs to manage their coordination and ensure proper connectivity and coverage. The MDRL model is trained using Proximal Policy Optimization (PPO) with Convolutional Neural Networks (CNNs) and a team-based reward. Conducted experiments show the superiority of the selection mechanism when compared to the benchmarks in terms of the percentages of selected UAVs, as well as the average QoUs and QoVs. The MDRL method proves efficient in producing UAV agents capable of maintaining coverage and connectivity and outperforms existing centralized benchmarks in terms of scalability. The cost analysis of the smart contract managing framework shows its feasibility and low complexity.

While the proposed framework demonstrates robust performance in UAV relay selection and coordination, there are limitations that could be addressed in future work. First, the framework does not consider the integration of charging stations or energy management mechanisms, which are critical for sustaining UAV operations over extended periods. Addressing this limitation would enhance the practicality of the system in real-world scenarios. Second, the simulations assume an idealized environment without obstacles or complexities that could affect communication between UAVs and vehicles, as well as their mobilities. In practice, factors such as physical obstructions, signal interference, and dynamic environmental conditions could impact the communication between the agents. Future work could extend the proposed methods to incorporate energy-aware strategies and account for environmental challenges to further improve the applicability and robustness of UAV-assisted IoV networks.

\bibliographystyle{model1-num-names}

\begin{thebibliography}{49}
\expandafter\ifx\csname natexlab\endcsname\relax\def\natexlab#1{#1}\fi
\providecommand{\bibinfo}[2]{#2}
\ifx\xfnm\relax \def\xfnm[#1]{\unskip,\space#1}\fi
\bibitem[{Hemmati et~al.(2023)Hemmati, Zarei, and Souri}]{HEMMATI2023200226}
\bibinfo{author}{A.~Hemmati}, \bibinfo{author}{M.~Zarei}, \bibinfo{author}{A.~Souri},
\newblock \bibinfo{title}{Uav-based internet of vehicles: A systematic literature review},
\newblock \bibinfo{journal}{Intelligent Systems with Applications} \bibinfo{volume}{18} (\bibinfo{year}{2023}) \bibinfo{pages}{200226}.
\bibitem[{Alagha et~al.(2021)Alagha, Mizouni, Singh, Otrok, and Ouali}]{alagha2021sdrs}
\bibinfo{author}{A.~Alagha}, \bibinfo{author}{R.~Mizouni}, \bibinfo{author}{S.~Singh}, \bibinfo{author}{H.~Otrok}, \bibinfo{author}{A.~Ouali},
\newblock \bibinfo{title}{Sdrs: A stable data-based recruitment system in iot crowdsensing for localization tasks},
\newblock \bibinfo{journal}{Journal of Network and Computer Applications} \bibinfo{volume}{177} (\bibinfo{year}{2021}) \bibinfo{pages}{102968}.
\bibitem[{Alagha et~al.(2020)Alagha, Singh, Otrok, and Mizouni}]{alagha2020rfls}
\bibinfo{author}{A.~Alagha}, \bibinfo{author}{S.~Singh}, \bibinfo{author}{H.~Otrok}, \bibinfo{author}{R.~Mizouni},
\newblock \bibinfo{title}{Rfls-resilient fault-proof localization system in iot and crowd-based sensing applications},
\newblock \bibinfo{journal}{Journal of Network and Computer Applications} \bibinfo{volume}{170} (\bibinfo{year}{2020}).
\bibitem[{Krishna(2020)}]{krishna2020survey}
\bibinfo{author}{M.~Krishna},
\newblock \bibinfo{title}{A survey uav-assisted vanet routing protocol},
\newblock \bibinfo{journal}{International Journal of Computer Science Trends and Technology (IJCST)--Vol} \bibinfo{volume}{8} (\bibinfo{year}{2020}).
\bibitem[{Tang et~al.(2020)Tang, Wei, Liu, Jiang, Wu, and Li}]{10.1007/978-981-15-9031-3_14}
\bibinfo{author}{C.~Tang}, \bibinfo{author}{X.~Wei}, \bibinfo{author}{C.~Liu}, \bibinfo{author}{H.~Jiang}, \bibinfo{author}{H.~Wu}, \bibinfo{author}{Q.~Li},
\newblock \bibinfo{title}{Uav-enabled social internet of vehicles: Roles, security issues and use cases},
\newblock in: \bibinfo{editor}{Y.~Xiang}, \bibinfo{editor}{Z.~Liu}, \bibinfo{editor}{J.~Li} (Eds.), \bibinfo{booktitle}{Security and Privacy in Social Networks and Big Data}, \bibinfo{publisher}{Springer Singapore}, \bibinfo{address}{Singapore}, \bibinfo{year}{2020}, pp. \bibinfo{pages}{153--163}.
\bibitem[{Su et~al.(2023)Su, Liwang, Chen, and Du}]{10117548}
\bibinfo{author}{Y.~Su}, \bibinfo{author}{M.~Liwang}, \bibinfo{author}{Z.~Chen}, \bibinfo{author}{X.~Du},
\newblock \bibinfo{title}{Toward optimal deployment of uav relays in uav-assisted internet of vehicles},
\newblock \bibinfo{journal}{IEEE Transactions on Vehicular Technology} \bibinfo{volume}{72} (\bibinfo{year}{2023}) \bibinfo{pages}{13392--13405}.
\bibitem[{Ng et~al.(2021)Ng, Lim, Dai, Xiong, Huang, Niyato, Hua, Leung, and Miao}]{9292475}
\bibinfo{author}{J.~S. Ng}, \bibinfo{author}{W.~Y.~B. Lim}, \bibinfo{author}{H.-N. Dai}, \bibinfo{author}{Z.~Xiong}, \bibinfo{author}{J.~Huang}, \bibinfo{author}{D.~Niyato}, \bibinfo{author}{X.-S. Hua}, \bibinfo{author}{C.~Leung}, \bibinfo{author}{C.~Miao},
\newblock \bibinfo{title}{Joint auction-coalition formation framework for communication-efficient federated learning in uav-enabled internet of vehicles},
\newblock \bibinfo{journal}{IEEE Transactions on Intelligent Transportation Systems} \bibinfo{volume}{22} (\bibinfo{year}{2021}) \bibinfo{pages}{2326--2344}.
\bibitem[{Abualola and Otrok(2021)}]{ABUALOLA2021100355}
\bibinfo{author}{H.~Abualola}, \bibinfo{author}{H.~Otrok},
\newblock \bibinfo{title}{Stable coalitions for urban-vanet: A hedonic game approach},
\newblock \bibinfo{journal}{Vehicular Communications} \bibinfo{volume}{30} (\bibinfo{year}{2021}) \bibinfo{pages}{100355}.
\bibitem[{Sami et~al.(2023)Sami, Saado, Saoudi, Mourad, Otrok, and Bentahar}]{10036008}
\bibinfo{author}{H.~Sami}, \bibinfo{author}{R.~Saado}, \bibinfo{author}{A.~E. Saoudi}, \bibinfo{author}{A.~Mourad}, \bibinfo{author}{H.~Otrok}, \bibinfo{author}{J.~Bentahar},
\newblock \bibinfo{title}{Opportunistic uav deployment for intelligent on-demand iov service management},
\newblock \bibinfo{journal}{IEEE Transactions on Network and Service Management} \bibinfo{volume}{20} (\bibinfo{year}{2023}) \bibinfo{pages}{3428--3442}.
\bibitem[{Abualola et~al.(2021)Abualola, Otrok, Barada, Al-Qutayri, and Al-Hammadi}]{ABUALOLA2021100290}
\bibinfo{author}{H.~Abualola}, \bibinfo{author}{H.~Otrok}, \bibinfo{author}{H.~Barada}, \bibinfo{author}{M.~Al-Qutayri}, \bibinfo{author}{Y.~Al-Hammadi},
\newblock \bibinfo{title}{Matching game theoretical model for stable relay selection in a uav-assisted internet of vehicles},
\newblock \bibinfo{journal}{Vehicular Communications} \bibinfo{volume}{27} (\bibinfo{year}{2021}) \bibinfo{pages}{100290}.
\bibitem[{Nakamoto(2009)}]{nsatoshi}
\bibinfo{author}{S.~Nakamoto},
\newblock \bibinfo{title}{Bitcoin: A peer-to-peer electronic cash system},
\newblock \bibinfo{journal}{Cryptography Mailing list at https://metzdowd.com}  (\bibinfo{year}{2009}).
\bibitem[{Kadadha and Otrok(2021)}]{KADADHA2021102502}
\bibinfo{author}{M.~Kadadha}, \bibinfo{author}{H.~Otrok},
\newblock \bibinfo{title}{A blockchain-enabled relay selection for qos-olsr in urban vanet: A stackelberg game model},
\newblock \bibinfo{journal}{Ad Hoc Networks} \bibinfo{volume}{117} (\bibinfo{year}{2021}) \bibinfo{pages}{102502}.
\bibitem[{Hammoud et~al.(2020)Hammoud, Sami, Mourad, Otrok, Mizouni, and Bentahar}]{9125437}
\bibinfo{author}{A.~Hammoud}, \bibinfo{author}{H.~Sami}, \bibinfo{author}{A.~Mourad}, \bibinfo{author}{H.~Otrok}, \bibinfo{author}{R.~Mizouni}, \bibinfo{author}{J.~Bentahar},
\newblock \bibinfo{title}{Ai, blockchain, and vehicular edge computing for smart and secure iov: Challenges and directions},
\newblock \bibinfo{journal}{IEEE Internet of Things Magazine} \bibinfo{volume}{3} (\bibinfo{year}{2020}) \bibinfo{pages}{68--73}.
\bibitem[{Sharma et~al.(2019)Sharma, Ghanshala, and Mohan}]{8911664}
\bibinfo{author}{S.~Sharma}, \bibinfo{author}{K.~K. Ghanshala}, \bibinfo{author}{S.~Mohan},
\newblock \bibinfo{title}{Blockchain-based internet of vehicles (iov): An efficient secure ad hoc vehicular networking architecture},
\newblock in: \bibinfo{booktitle}{2019 IEEE 2nd 5G World Forum (5GWF)}, pp. \bibinfo{pages}{452--457}.
\bibitem[{Tu et~al.(2023)Tu, Yu, Badshah, Waqas, Halim, and Ahmad}]{10104127}
\bibinfo{author}{S.~Tu}, \bibinfo{author}{H.~Yu}, \bibinfo{author}{A.~Badshah}, \bibinfo{author}{M.~Waqas}, \bibinfo{author}{Z.~Halim}, \bibinfo{author}{I.~Ahmad},
\newblock \bibinfo{title}{Secure internet of vehicles (iov) with decentralized consensus blockchain mechanism},
\newblock \bibinfo{journal}{IEEE Transactions on Vehicular Technology} \bibinfo{volume}{72} (\bibinfo{year}{2023}) \bibinfo{pages}{11227--11236}.
\bibitem[{Alladi et~al.(2020)Alladi, Chamola, Sahu, and Guizani}]{ALLADI2020100249}
\bibinfo{author}{T.~Alladi}, \bibinfo{author}{V.~Chamola}, \bibinfo{author}{N.~Sahu}, \bibinfo{author}{M.~Guizani},
\newblock \bibinfo{title}{Applications of blockchain in unmanned aerial vehicles: A review},
\newblock \bibinfo{journal}{Vehicular Communications} \bibinfo{volume}{23} (\bibinfo{year}{2020}) \bibinfo{pages}{100249}.
\bibitem[{Islam et~al.(2022)Islam, Khan, Saad, Tariq, and Kim}]{islam2022dynamic}
\bibinfo{author}{M.~M. Islam}, \bibinfo{author}{M.~T.~R. Khan}, \bibinfo{author}{M.~M. Saad}, \bibinfo{author}{M.~A. Tariq}, \bibinfo{author}{D.~Kim},
\newblock \bibinfo{title}{Dynamic positioning of uavs to improve network coverage in vanets},
\newblock \bibinfo{journal}{Vehicular Communications} \bibinfo{volume}{36} (\bibinfo{year}{2022}) \bibinfo{pages}{100498}.
\bibitem[{Sedjelmaci et~al.(2019)Sedjelmaci, Messous, Senouci, and Brahmi}]{sedjelmaci2019toward}
\bibinfo{author}{H.~Sedjelmaci}, \bibinfo{author}{M.~A. Messous}, \bibinfo{author}{S.~M. Senouci}, \bibinfo{author}{I.~H. Brahmi},
\newblock \bibinfo{title}{Toward a lightweight and efficient uav-aided vanet},
\newblock \bibinfo{journal}{Transactions on Emerging Telecommunications Technologies} \bibinfo{volume}{30} (\bibinfo{year}{2019}) \bibinfo{pages}{e3520}.
\bibitem[{Raza et~al.(2021)Raza, Bukhari, Aadil, and Iqbal}]{raza2021uav}
\bibinfo{author}{A.~Raza}, \bibinfo{author}{S.~H.~R. Bukhari}, \bibinfo{author}{F.~Aadil}, \bibinfo{author}{Z.~Iqbal},
\newblock \bibinfo{title}{An uav-assisted vanet architecture for intelligent transportation system in smart cities},
\newblock \bibinfo{journal}{International Journal of Distributed Sensor Networks} \bibinfo{volume}{17} (\bibinfo{year}{2021}) \bibinfo{pages}{15501477211031750}.
\bibitem[{Samir et~al.(2021)Samir, Ebrahimi, Assi, Sharafeddine, and Ghrayeb}]{9082162}
\bibinfo{author}{M.~Samir}, \bibinfo{author}{D.~Ebrahimi}, \bibinfo{author}{C.~Assi}, \bibinfo{author}{S.~Sharafeddine}, \bibinfo{author}{A.~Ghrayeb},
\newblock \bibinfo{title}{Leveraging uavs for coverage in cell-free vehicular networks: A deep reinforcement learning approach},
\newblock \bibinfo{journal}{IEEE Transactions on Mobile Computing} \bibinfo{volume}{20} (\bibinfo{year}{2021}) \bibinfo{pages}{2835--2847}.
\bibitem[{Yuan et~al.(2021)Yuan, Rothenberg, Obraczka, Barakat, and Turletti}]{9585312}
\bibinfo{author}{T.~Yuan}, \bibinfo{author}{C.~E. Rothenberg}, \bibinfo{author}{K.~Obraczka}, \bibinfo{author}{C.~Barakat}, \bibinfo{author}{T.~Turletti},
\newblock \bibinfo{title}{Harnessing uavs for fair 5g bandwidth allocation in vehicular communication via deep reinforcement learning},
\newblock \bibinfo{journal}{IEEE Transactions on Network and Service Management} \bibinfo{volume}{18} (\bibinfo{year}{2021}) \bibinfo{pages}{4063--4074}.
\bibitem[{Oubbati et~al.(2021)Oubbati, Atiquzzaman, Baz, Alhakami, and Ben-Othman}]{9566766}
\bibinfo{author}{O.~S. Oubbati}, \bibinfo{author}{M.~Atiquzzaman}, \bibinfo{author}{A.~Baz}, \bibinfo{author}{H.~Alhakami}, \bibinfo{author}{J.~Ben-Othman},
\newblock \bibinfo{title}{Dispatch of uavs for urban vehicular networks: A deep reinforcement learning approach},
\newblock \bibinfo{journal}{IEEE Transactions on Vehicular Technology} \bibinfo{volume}{70} (\bibinfo{year}{2021}) \bibinfo{pages}{13174--13189}.
\bibitem[{Kadadha et~al.(2024)Kadadha, Mizouni, Singh, Otrok, and Mourad}]{KADADHA2024100761}
\bibinfo{author}{M.~Kadadha}, \bibinfo{author}{R.~Mizouni}, \bibinfo{author}{S.~Singh}, \bibinfo{author}{H.~Otrok}, \bibinfo{author}{A.~Mourad},
\newblock \bibinfo{title}{Crowdsourced vehicles and uavs for last-mile delivery application using blockchain-hosted matching mechanism},
\newblock \bibinfo{journal}{Vehicular Communications} \bibinfo{volume}{47} (\bibinfo{year}{2024}) \bibinfo{pages}{100761}.
\bibitem[{Kadadha et~al.(2022)Kadadha, Singh, Mizouni, and Otrok}]{kadadha2022context}
\bibinfo{author}{M.~Kadadha}, \bibinfo{author}{S.~Singh}, \bibinfo{author}{R.~Mizouni}, \bibinfo{author}{H.~Otrok},
\newblock \bibinfo{title}{A context-aware blockchain-based crowdsourcing framework: Open challenges and opportunities},
\newblock \bibinfo{journal}{IEEE Access}  (\bibinfo{year}{2022}).
\bibitem[{Samir et~al.(2019)Samir, Sharafeddine, Assi, Nguyen, and Ghrayeb}]{8716508}
\bibinfo{author}{M.~Samir}, \bibinfo{author}{S.~Sharafeddine}, \bibinfo{author}{C.~Assi}, \bibinfo{author}{T.~M. Nguyen}, \bibinfo{author}{A.~Ghrayeb},
\newblock \bibinfo{title}{Trajectory planning and resource allocation of multiple uavs for data delivery in vehicular networks},
\newblock \bibinfo{journal}{IEEE Networking Letters} \bibinfo{volume}{1} (\bibinfo{year}{2019}) \bibinfo{pages}{107--110}.
\bibitem[{Lin et~al.(2020)Lin, Fu, Zhao, Min, Al-Dubai, and Gacanin}]{9076813}
\bibinfo{author}{N.~Lin}, \bibinfo{author}{L.~Fu}, \bibinfo{author}{L.~Zhao}, \bibinfo{author}{G.~Min}, \bibinfo{author}{A.~Al-Dubai}, \bibinfo{author}{H.~Gacanin},
\newblock \bibinfo{title}{A novel multimodal collaborative drone-assisted vanet networking model},
\newblock \bibinfo{journal}{IEEE Transactions on Wireless Communications} \bibinfo{volume}{19} (\bibinfo{year}{2020}) \bibinfo{pages}{4919--4933}.
\bibitem[{Hadiwardoyo et~al.(2020)Hadiwardoyo, Calafate, Cano, Krinkin, Klionskiy, Hern{\'a}ndez-Orallo, and Manzoni}]{hadiwardoyo2020three}
\bibinfo{author}{S.~A. Hadiwardoyo}, \bibinfo{author}{C.~T. Calafate}, \bibinfo{author}{J.-C. Cano}, \bibinfo{author}{K.~Krinkin}, \bibinfo{author}{D.~Klionskiy}, \bibinfo{author}{E.~Hern{\'a}ndez-Orallo}, \bibinfo{author}{P.~Manzoni},
\newblock \bibinfo{title}{Three dimensional uav positioning for dynamic uav-to-car communications},
\newblock \bibinfo{journal}{Sensors} \bibinfo{volume}{20} (\bibinfo{year}{2020}) \bibinfo{pages}{356}.
\bibitem[{Ahmed et~al.(2021)Ahmed, Sheltami, Mahmoud, Imran, and Shoaib}]{9400369}
\bibinfo{author}{G.~A. Ahmed}, \bibinfo{author}{T.~R. Sheltami}, \bibinfo{author}{A.~S. Mahmoud}, \bibinfo{author}{M.~Imran}, \bibinfo{author}{M.~Shoaib},
\newblock \bibinfo{title}{A novel collaborative iod-assisted vanet approach for coverage area maximization},
\newblock \bibinfo{journal}{IEEE Access} \bibinfo{volume}{9} (\bibinfo{year}{2021}) \bibinfo{pages}{61211--61223}.
\bibitem[{Noh et~al.(2020)Noh, Jeon, and Chae}]{9044857}
\bibinfo{author}{S.-C. Noh}, \bibinfo{author}{H.-B. Jeon}, \bibinfo{author}{C.-B. Chae},
\newblock \bibinfo{title}{Energy-efficient deployment of multiple uavs using ellipse clustering to establish base stations},
\newblock \bibinfo{journal}{IEEE Wireless Communications Letters} \bibinfo{volume}{9} (\bibinfo{year}{2020}) \bibinfo{pages}{1155--1159}.
\bibitem[{Samir et~al.(2020)Samir, Ebrahimi, Assi, Sharafeddine, and Ghrayeb}]{8960481}
\bibinfo{author}{M.~Samir}, \bibinfo{author}{D.~Ebrahimi}, \bibinfo{author}{C.~Assi}, \bibinfo{author}{S.~Sharafeddine}, \bibinfo{author}{A.~Ghrayeb},
\newblock \bibinfo{title}{Trajectory planning of multiple dronecells in vehicular networks: A reinforcement learning approach},
\newblock \bibinfo{journal}{IEEE Networking Letters} \bibinfo{volume}{2} (\bibinfo{year}{2020}) \bibinfo{pages}{14--18}.
\bibitem[{Damani et~al.(2021)Damani, Luo, Wenzel, and Sartoretti}]{damani2021primal}
\bibinfo{author}{M.~Damani}, \bibinfo{author}{Z.~Luo}, \bibinfo{author}{E.~Wenzel}, \bibinfo{author}{G.~Sartoretti},
\newblock \bibinfo{title}{Primal $ \_2 $: Pathfinding via reinforcement and imitation multi-agent learning-lifelong},
\newblock \bibinfo{journal}{IEEE Robotics and Automation Letters} \bibinfo{volume}{6} (\bibinfo{year}{2021}) \bibinfo{pages}{2666--2673}.
\bibitem[{Bayerlein et~al.(2021)Bayerlein, Theile, Caccamo, and Gesbert}]{bayerlein2021multi}
\bibinfo{author}{H.~Bayerlein}, \bibinfo{author}{M.~Theile}, \bibinfo{author}{M.~Caccamo}, \bibinfo{author}{D.~Gesbert},
\newblock \bibinfo{title}{Multi-uav path planning for wireless data harvesting with deep reinforcement learning},
\newblock \bibinfo{journal}{IEEE Open Journal of the Communications Society} \bibinfo{volume}{2} (\bibinfo{year}{2021}) \bibinfo{pages}{1171--1187}.
\bibitem[{Alagha et~al.(2023)Alagha, Mizouni, Bentahar, Otrok, and Singh}]{alagha2023multi}
\bibinfo{author}{A.~Alagha}, \bibinfo{author}{R.~Mizouni}, \bibinfo{author}{J.~Bentahar}, \bibinfo{author}{H.~Otrok}, \bibinfo{author}{S.~Singh},
\newblock \bibinfo{title}{Multi-agent deep reinforcement learning with demonstration cloning for target localization},
\newblock \bibinfo{journal}{IEEE Internet of Things Journal}  (\bibinfo{year}{2023}).
\bibitem[{Lee and Kim(2023)}]{lee2023multi}
\bibinfo{author}{W.~Lee}, \bibinfo{author}{T.~Kim},
\newblock \bibinfo{title}{Multi-agent reinforcement learning in controlling offloading ratio and trajectory for multi-uav mobile edge computing},
\newblock \bibinfo{journal}{IEEE Internet of Things Journal}  (\bibinfo{year}{2023}).
\bibitem[{Jiang et~al.(2021)Jiang, Givigi, and Delamer}]{jiang2021marl}
\bibinfo{author}{B.~Jiang}, \bibinfo{author}{S.~N. Givigi}, \bibinfo{author}{J.-A. Delamer},
\newblock \bibinfo{title}{A marl approach for optimizing positions of vanet aerial base-stations on a sparse highway},
\newblock \bibinfo{journal}{IEEE Access} \bibinfo{volume}{9} (\bibinfo{year}{2021}) \bibinfo{pages}{133989--134004}.
\bibitem[{quo(2018{\natexlab{a}})}]{quorum}
\bibinfo{title}{Quorum: a permissioned implementation of ethereum supporting data privacy}, \bibinfo{year}{2018}{\natexlab{a}}. \bibinfo{note}{Accessed: 27-09-2023}.
\bibitem[{quo(2018{\natexlab{b}})}]{quorum_performance}
\bibinfo{title}{Performance evaluation of the quorum blockchain platform},
\newblock \bibinfo{journal}{arXiv:1809.03421}  (\bibinfo{year}{2018}{\natexlab{b}}). \bibinfo{note}{Accessed: 2024-11-21}.
\bibitem[{Gronauer and Diepold(2021)}]{gronauer2021multi}
\bibinfo{author}{S.~Gronauer}, \bibinfo{author}{K.~Diepold},
\newblock \bibinfo{title}{Multi-agent deep reinforcement learning: a survey},
\newblock \bibinfo{journal}{Artificial Intelligence Review}  (\bibinfo{year}{2021}) \bibinfo{pages}{1--49}.
\bibitem[{Alagha et~al.(2023)Alagha, Bentahar, Otrok, Singh, and Mizouni}]{alagha2023blockchain}
\bibinfo{author}{A.~Alagha}, \bibinfo{author}{J.~Bentahar}, \bibinfo{author}{H.~Otrok}, \bibinfo{author}{S.~Singh}, \bibinfo{author}{R.~Mizouni},
\newblock \bibinfo{title}{Blockchain-assisted demonstration cloning for multi-agent deep reinforcement learning},
\newblock \bibinfo{journal}{IEEE Internet of Things Journal}  (\bibinfo{year}{2023}).
\bibitem[{Alagha et~al.(2024)Alagha, Otrok, Singh, Mizouni, and Bentahar}]{alagha4753209blockchain}
\bibinfo{author}{A.~Alagha}, \bibinfo{author}{H.~Otrok}, \bibinfo{author}{S.~Singh}, \bibinfo{author}{R.~Mizouni}, \bibinfo{author}{J.~Bentahar},
\newblock \bibinfo{title}{Blockchain-based crowdsourced deep reinforcement learning as a service},
\newblock \bibinfo{journal}{Available at SSRN 4753209}  (\bibinfo{year}{2024}).
\bibitem[{Schulman et~al.(2017)}]{schulman2017proximal}
\bibinfo{author}{J.~Schulman}, et~al.,
\newblock \bibinfo{title}{Proximal policy optimization algorithms},
\newblock \bibinfo{journal}{arXiv preprint arXiv:1707.06347}  (\bibinfo{year}{2017}).
\bibitem[{Baker et~al.(2020)Baker, Kanitscheider, Markov, Wu, Powell, McGrew, and Mordatch}]{baker2019emergent}
\bibinfo{author}{B.~Baker}, \bibinfo{author}{I.~Kanitscheider}, \bibinfo{author}{T.~Markov}, \bibinfo{author}{Y.~Wu}, \bibinfo{author}{G.~Powell}, \bibinfo{author}{B.~McGrew}, \bibinfo{author}{I.~Mordatch},
\newblock \bibinfo{title}{Emergent tool use from multi-agent autocurricula},
\newblock in: \bibinfo{booktitle}{2020 Proc. Int. Conf. on Learning Representations (ICLR)}.
\bibitem[{Alagha et~al.(2022)Alagha, Singh, Mizouni, Bentahar, and Otrok}]{alagha2022target}
\bibinfo{author}{A.~Alagha}, \bibinfo{author}{S.~Singh}, \bibinfo{author}{R.~Mizouni}, \bibinfo{author}{J.~Bentahar}, \bibinfo{author}{H.~Otrok},
\newblock \bibinfo{title}{Target localization using multi-agent deep reinforcement learning with proximal policy optimization},
\newblock \bibinfo{journal}{Future Generation Computer Systems} \bibinfo{volume}{136} (\bibinfo{year}{2022}) \bibinfo{pages}{342--357}.
\bibitem[{Alagha et~al.(2024)Alagha, Mizouni, Singh, Bentahar, and Otrok}]{alagha4872731adaptive}
\bibinfo{author}{A.~Alagha}, \bibinfo{author}{R.~Mizouni}, \bibinfo{author}{S.~Singh}, \bibinfo{author}{J.~Bentahar}, \bibinfo{author}{H.~Otrok},
\newblock \bibinfo{title}{Adaptive target localization under uncertainty using multi-agent deep reinforcement learning with knowledge transfer},
\newblock \bibinfo{journal}{Available at SSRN 4872731}  (\bibinfo{year}{2024}).
\bibitem[{LeCun et~al.(2015)}]{lecun2015lenet}
\bibinfo{author}{Y.~LeCun}, et~al.,
\newblock \bibinfo{title}{Lenet-5, convolutional neural networks},
\newblock \bibinfo{journal}{URL: http://yann. lecun. com/exdb/lenet} \bibinfo{volume}{20} (\bibinfo{year}{2015}) \bibinfo{pages}{14}.
\bibitem[{Fan et~al.(2022)Fan, Wu, Liao, Cao, Guo, Sartoretti, and Wu}]{fan2022deep}
\bibinfo{author}{M.~Fan}, \bibinfo{author}{Y.~Wu}, \bibinfo{author}{T.~Liao}, \bibinfo{author}{Z.~Cao}, \bibinfo{author}{H.~Guo}, \bibinfo{author}{G.~Sartoretti}, \bibinfo{author}{G.~Wu},
\newblock \bibinfo{title}{Deep reinforcement learning for uav routing in the presence of multiple charging stations},
\newblock \bibinfo{journal}{IEEE Transactions on Vehicular Technology} \bibinfo{volume}{72} (\bibinfo{year}{2022}) \bibinfo{pages}{5732--5746}.
\bibitem[{Liu et~al.(2020)Liu, Piao, and Tang}]{liu2020energy}
\bibinfo{author}{C.~H. Liu}, \bibinfo{author}{C.~Piao}, \bibinfo{author}{J.~Tang},
\newblock \bibinfo{title}{Energy-efficient uav crowdsensing with multiple charging stations by deep learning},
\newblock in: \bibinfo{booktitle}{IEEE INFOCOm 2020-IEEE conference on computer communications}, \bibinfo{organization}{IEEE}, pp. \bibinfo{pages}{199--208}.
\bibitem[{Dehejia and Wahba(2002)}]{dehejia2002propensity}
\bibinfo{author}{R.~H. Dehejia}, \bibinfo{author}{S.~Wahba},
\newblock \bibinfo{title}{Propensity score-matching methods for nonexperimental causal studies},
\newblock \bibinfo{journal}{Review of Economics and statistics} \bibinfo{volume}{84} (\bibinfo{year}{2002}) \bibinfo{pages}{151--161}.
\bibitem[{Sutton and Barto(2018)}]{sutton2018reinforcement}
\bibinfo{author}{R.~S. Sutton}, \bibinfo{author}{A.~G. Barto}, \bibinfo{title}{Reinforcement learning: An introduction}, \bibinfo{publisher}{MIT press}, \bibinfo{year}{2018}.

\end{thebibliography}

\end{document}